\documentclass[useAMS,usenatbib,fleqn,twocolumn]{aastex7}
\usepackage{multirow}
\usepackage{graphicx}
\usepackage{xcolor,soul}
\usepackage{amssymb,amsmath}
\usepackage{xspace}
\usepackage{tabularx} 
\usepackage[T1]{fontenc} 

\def\sgra{Sgr~A$^*$\xspace}
\def\m87{M87$^*$\xspace}

\begin{document}

\shorttitle{QU Loop Speeds}
\title{Dynamical Inference from Polarized Light Curves of Sagittarius A$^*$}

\shortauthors{Ricarte et al.}
\correspondingauthor{Angelo Ricarte}
\email{angelo.ricarte@cfa.harvard.edu}

\author[0000-0001-5287-0452]{Angelo Ricarte}
\affiliation{Black Hole Initiative at Harvard University, 20 Garden Street, Cambridge, MA 02138, USA}
\affiliation{Center for Astrophysics | Harvard \& Smithsonian, 60 Garden Street, Cambridge, MA 02138, USA}
\email{angelo.ricarte@cfa.harvard.edu}

\author[0000-0003-2886-2377]{Nicholas S. Conroy}
\affiliation{Department of Astronomy, University of Illinois at Urbana-Champaign, 1002 West Green Street, Urbana, IL 61801, USA}
\email{nconroy2@illinois.edu}

\author[0000-0002-8635-4242]{Maciek Wielgus}
\affiliation{Instituto de Astrofísica de Andalucía-CSIC, Glorieta de la Astronomía s/n, E-18008 Granada, Spain}
\email{maciek.wielgus@gmail.com}

\author[0000-0002-7179-3816]{Daniel Palumbo}
\affiliation{Black Hole Initiative at Harvard University, 20 Garden Street, Cambridge, MA 02138, USA}
\affiliation{Center for Astrophysics | Harvard \& Smithsonian, 60 Garden Street, Cambridge, MA 02138, USA}
\email{daniel.palumbo@cfa.harvard.edu }

\author[0000-0002-2791-5011]{Razieh Emami}
\affiliation{Center for Astrophysics | Harvard \& Smithsonian, 60 Garden Street, Cambridge, MA 02138, USA}
\email{razieh.emami_meibody@cfa.harvard.edu}

\author[0000-0001-6337-6126]{Chi-kwan Chan}
\affiliation{Steward Observatory and Department of Astronomy, University of Arizona, 933 N. Cherry Avenue, Tucson, AZ 85721, USA}
\affiliation{Data Science Institute, University of Arizona, 1230 N. Cherry Avenue, Tucson, AZ 85721, USA}
\affiliation{Program in Applied Mathematics, University of Arizona, 617 N. Santa Rita, Tucson, AZ 85721, USA}
\email{chanc@arizona.edu}

\date{\today}

\begin{abstract}

Polarimetric light curves of Sagittarius A$^*$ (\sgra) sometimes exhibit loops in the Stokes $Q$ and $U$ plane over time, often interpreted as orbiting hotspot motion.  In this work, we apply the differential geometry of planar curves to develop a new technique for estimating polarimetric rotation rates.  Applying this technique to 230 GHz light curves of \sgra, we find evidence of clockwise motion not only during a post-flare period on 2017 April 11th, as previously discovered, but also during the quiescent days imaged by the Event Horizon Telescope (EHT).  The data exhibit a clockwise fraction of $0.65 \pm 0.09$ and an overall $Q-U$ rotation rate of $-2.6 \pm 0.6 \ \mathrm{deg}\,t_g^{-1}$.  We analyze a library of General Relativistic Magnetohydrodynamic (GRMHD) simulations and find that face-on, clockwise-rotating models with strong magnetic fields are most likely to be consistent with the observations. These results are consistent with EHT and GRAVITY Collaboration studies, and indirectly support an interpretation in which the polarized image of \sgra has been rotated by an external Faraday screen. This technique offers a novel probe of event horizon scale dynamics that complements dynamical reconstructions.

\end{abstract}


\section{Introduction}
\label{sec:introduction}

Our Galaxy hosts a central radio source known as Sagittarius A$^*$ \citep[\sgra;][]{Balick&Brown1974}, generally agreed to be a $4 \times 10^6 \ M_\odot$ supermassive black hole (BH), whose mass has been constrained by both stellar orbits \citep{Schoedel+2002,Ghez+2003,Ghez+2008,Gillessen+2017,Do+2019,Gravity+2022} and direct imaging \citep{EHTC+2022a}. The first full-Stokes images of \sgra have recently been published by the Event Horizon Telescope (EHT) collaboration, revealing a rotationally symmetric ring and linear polarization pattern \citep{EHTC+2022d,EHTC+2024b}. The image of \sgra and its multi-wavelength properties are consistent with a hot accretion flow \citep[e.g.,][]{Yuan&Narayan2014} with dynamically important magnetic fields \citep{Gravity+2020b,EHTC+2022e,EHTC+2024c}.  

At millimeter wavelengths, \sgra exhibits time-variability on timescales from minutes to years \citep{Bower+2005,Marrone2006,Dexter+2014,Bower+2018,Wielgus+2022,Wielgus+2024}.  For the reconstruction of the EHT's first static image of \sgra, structural variability was evident in the measured interferometric visibilities, necessitating the development of mitigation techniques during image reconstruction \citep{Georgiev+2022,Broderick+2022,EHTC+2022d}. This implies that horizon scale dynamics are accessible by the EHT, from which dynamical reconstructions associated with orbital motion can be extracted \citep[e.g.,][]{Knollmueller+2023,Levis2024}.

Recent numerical simulations of EHT targets highlight polarization as a sensitive probe of accretion flow and space-time properties including BH spin and magnetic field state \citep[e.g.,][]{Moscibrodzka+2017,Palumbo+2020,Tsunetoe+2021,Narayan+2021,EHTC+2021b,Ricarte+2023c,Emami+2023d,Qiu+2023,EHTC+2023,Chael+2023,EHTC+2024c}.  In addition to spatially resolved motion, polarized temporal evolution offers an alternative probe into horizon scale dynamics. Loops in the Stokes $Q$ and $U$ plane over time, or ``$Q-U$ loops,'' have been observed at both millimeter \citep{Marrone+2006,Wielgus2022QU} and near infra-red (NIR) wavelengths \citep{Gravity2018QU,Gravity+2023,Yfantis+2024b}.  Four of the six NIR $Q-U$ loops observed by the GRAVITY collaboration are associated with clockwise centroid motion, all measured using the Very Large Telescope Interferometer \citep[VLTI;][]{Gravity+2023}.  The salient features of $Q-U$ loops can be naturally explained using hotspot models, wherein a spot of emission selectively illuminates regions within a rotationally symmetric magnetic field as it orbits the BH \citep{Broderick&Loeb2006,Hamaus+2009,Gelles+2021,Vos2022,Vincent2024,Yfantis2024}.  Previous works propose physical connections between hotspots and flaring behavior, which temporarily excite electrons to high energy. This hotspot can then take the form of a flux tube, or plasmoid, possibly following a flux eruption event \citep{Dexter+2020,Chatterjee+2021,Jia+2023,Ripperda+2022,Aimar+2023,ElMellah+2023,Najafi-Ziyazi+2024,Grigorian&Dexter2024,Antonopoulou+2025}.  Observationally, a connection between flares and $Q-U$ loops was established by \citet{Wielgus2022QU}, where a prominent polarimetric loop appeared following an X-ray flare detected by {\it Chandra}.

Studies of $Q-U$ loops have been mostly focused on flares, or the most visually obvious loops in the $Q-U$ plane.  However, the growing volume of polarized light curves of \sgra, with exquisite signal-to-noise enabled by the Atacama Large Millimeter Array (ALMA), motivates the development of tools to study patterns in the $Q-U$ plane in a more generic way, including quiescent periods.  In addition, ongoing and planned dynamical reconstructions of both \sgra and M$\,87^*$ motivate theoretical studies connecting $Q-U$ loops and Stokes $I$ pattern speeds \citep{Conroy+2023}.

The outline of the paper is as follows.  In \autoref{sec:methodology}, we briefly summarize the simulation library we use for analysis and a novel method for computing average rotation rates in the $Q-U$ plane.  In \autoref{sec:light_curves}, we apply our new method to polarized \sgra light curves taken using ALMA during the 2017 EHT campaign, obtaining consistent indications of clockwise motion on these days.  In \autoref{sec:model_comparison}, we compare the data with our simulation library, obtaining significant constraints that favor clockwise-rotating accretion flows.  We discuss connections with other observing frequencies and alternative dynamical tracers in \autoref{sec:discussion}.  Our results are summarized in \autoref{sec:conclusion}.

\section{Methodology}
\label{sec:methodology}

We compute Stokes $Q$ and $U$ light curves from General Relativistic Magnetohydrodynamics (GRMHD) models of \sgra.  Then, we apply the differential geometry of planar curves to define an average rotation rate for each model.

\subsection{Images from GRMHD}
\label{sec:images}

We use as our starting point a library of polarized model images of \sgra presented in \citet{EHTC+2022e,EHTC+2024c}, following the {\sc Patoka} simulation pipeline \citep{Wong+2022}.  Specifically, we consider the library run using the General Relativistic Magnetohydrodynamics (GRMHD) code KHARMA \citep{Prather2024}, subsequently ray-traced using the General Relativistic Ray-tracing (GRRT) code {\sc ipole} \citep{Moscibrodzka&Gammie2018}. GRRT is performed assuming relativistic thermal electron distribution functions.  We refer readers to \citet{Dhruv+2025} for additional details. 

This library includes 2 magnetic field states (strongly magnetized Magnetically Arrested Disk or ``MAD'' \citep{Bisnovatyi-Kogan&Ruzmaikin1974,Igumenshchev+2003,Narayan+2003,Tchekhovskoy+2011}  models, and more weakly magnetized Standard and Normal Evolution or ``SANE'' \citep{DeVilliers+2003,Gammie+2003,Narayan+2012,Sadowski+2013} models), 5 BH spins ($a_\bullet \in \{0,\pm0.5,\pm0.94\}$), 9 inclinations ($i \in \{10^\circ,30^\circ,...170^\circ\}$), 4 $R_\mathrm{high}$ values ($R_\mathrm{high} \in \{1,10,40,160\}$), and 2 magnetic field polarities (either ``aligned'' or ``reversed'' with respect to the disk angular momentum), for a total of 720 parameter combinations.  In this work we only include the subset with ``aligned'' magnetic fields, which were ray-traced at a higher cadence ($5 \ t_g$ rather than only $30 \ t_g$, where $r_g \equiv GMc^{-2}$ and $t_g\equiv r_g/c \approx 20 \ \mathrm{s}$ for \sgra).  Flipping the magnetic field polarity is expected to predominantly affect circular polarization and impart an overall electric vector position angle (EVPA) shift \citep[e.g.,][]{Emami+2023,Qiu+2023}, neither of which would affect the metrics in consideration.  

In our conventions, $a_\bullet < 0$ corresponds to a retrograde accretion flow, where the BH and disk angular momentum vectors are anti-aligned, and $a_\bullet > 0$ corresponds to a prograde accretion flow.  Inclination is defined relative to the spin vector of the accretion disk, not that of the BH, such that $i>90^\circ$ models have accretion disks that rotate clockwise on the sky, while $i<90^\circ$ models rotate counter-clockwise.  

Because ions and electrons are not expected to be in thermal equilibrium, the parameter $R_\mathrm{high}$ modulates the ion-to-electron temperature ratio via the \citet{Moscibrodzka+2016} prescription,

\begin{equation}
    \frac{T_i}{T_e} = \frac{1}{1+\beta^2} + R_\mathrm{high}\frac{\beta^2}{1+\beta^2}. \label{eqn:r_beta}
\end{equation}

Each simulation was initialized with a \citet{Fishbone&Moncrief1976} torus initial condition and an adiabatic index of $4/3$, then run for a total duration of 30000 $t_g$.  We analyze the quasi-steady state $15000-30000 \ t_g$ period of these simulations, during which the inner accretion flow has reached inflow equilibrium and no obvious signatures of relaxation from the initial conditions remain.

We refer readers to Appendix A of \citet{EHTC+2024c} for an overview of the observational impacts of changing each of these parameters on the resultant polarized images.  EHT theoretical studies so far have identified a ``best bet'' model within this simulation set that we will refer to in this paper.  It has the following parameters:  MAD, $a_\bullet=0.94$, $i=150^\circ$, $R_\mathrm{high}=160$, aligned fields.  This model passes most full-Stokes and multi-wavelength constraints considered in \citet{EHTC+2022e,EHTC+2024c}, but like most simulations in these studies, it is more variable in Stokes $I$ than the observational data.

\subsection{Differential Geometry of Loops}
\label{sec:loops}

\begin{figure*}
    \centering
    \includegraphics[width=\textwidth]{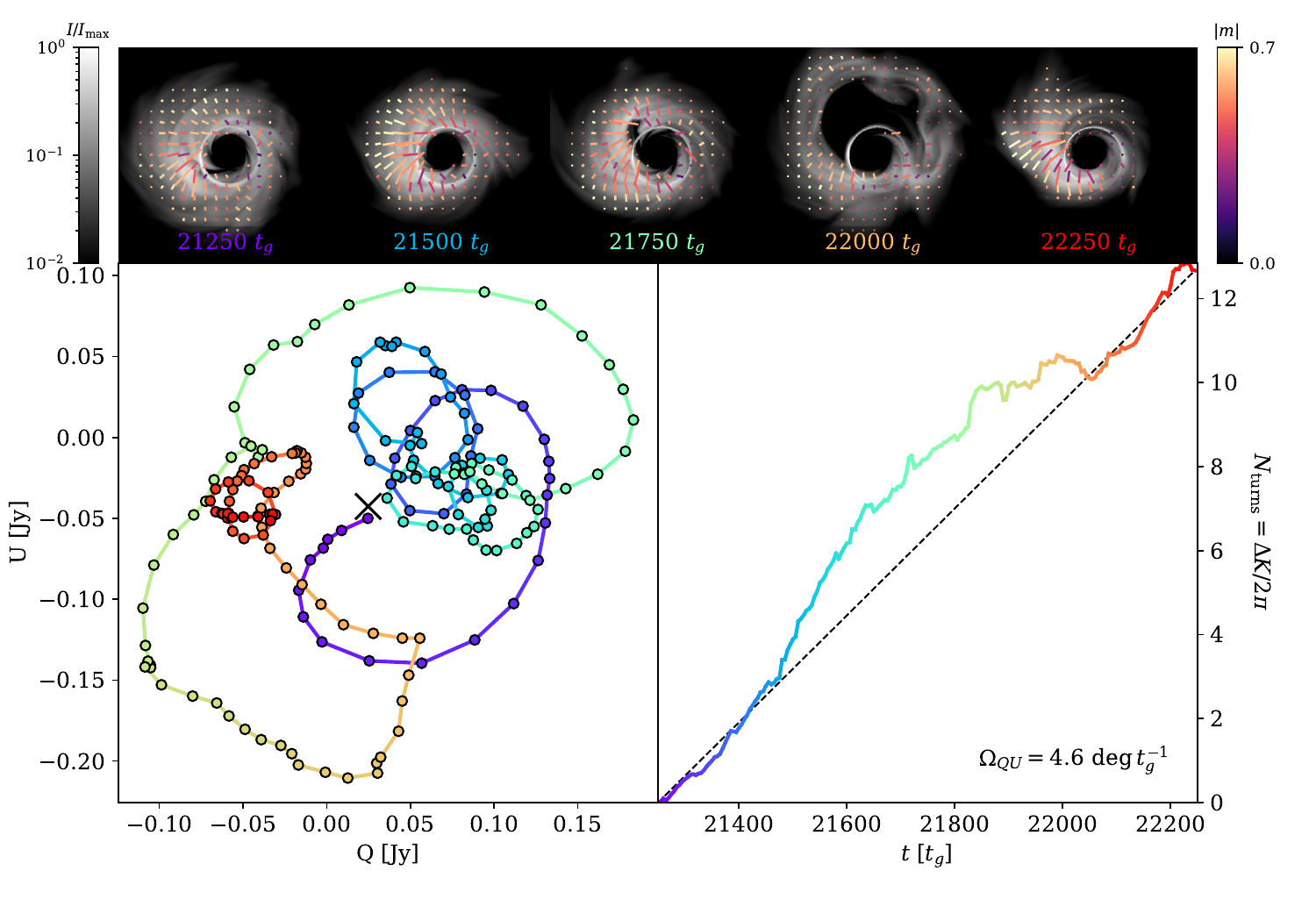}
    \caption{{\it Top:} Equally spaced snapshots of the MAD, $a_\bullet=0.5$, $R_\mathrm{high}=1$, $i=30^\circ$ model, visualized in log-scale. Tick lengths encode total linear polarization and colors encode the linear polarization fraction. {\it Bottom left:} $Q$ and $U$ plotted as a function of time, where an ``X'' marks the mean.  While the model makes many loops during this time period, it does not close a single loop around the mean value. Two separate flux eruption events between roughly 21,750 and 22,000 $t_g$ are associated with a wide arc on this plane. {\it Bottom right: } Number of turns made by the model over time following \autoref{eqn:total_curvature}.  As illustrated here, $\Omega_{QU}$ is estimated from the average slope of this curve.}
    \label{fig:example}
\end{figure*}

In the $Q-U$ plane, we use the differential geometry of planar curves to compute the signed curvature of a curve parameterized by $Q(s(t))$ and $U(s(t))$ \citep{Millman&Parker1977}.  Here, $s(t)$ is the arc length of the curve, computable by integrating 

\begin{equation}
    \frac{ds}{dt} = \sqrt{\left( \frac{dQ}{dt} \right) ^2+\left(\frac{dU}{dt}\right)^2}. \label{eqn:arclength}
\end{equation}

\noindent Then, the local curvature is given by 

\begin{equation}
    k(s) = \frac{\dot{Q}\ddot{U}-\dot{U}\ddot{Q}}{(\dot{Q}^2+\dot{U}^2)^{3/2}}. \label{eqn:local_curvature}
\end{equation}

\noindent where a $\cdot$ denotes a derivative with respect to $s$.  Note that $k$ is signed, returning positive values for counter-clockwise curves and negative values for clockwise curves.  It is related to the radius of curvature $r_c$ via

\begin{equation}
    r_c(s) = \frac{1}{|k(s)|}. \label{eqn:radius_of_curvature}
\end{equation}

The total curvature for a curve segment (in radians) parameterized from the start of the segment to arclength $s$ is given by

\begin{align}
    K(s) = \int_{0}^{s} k(s')ds', \label{eqn:total_curvature_s}
\end{align}

\noindent or equivalently from time 0 to $t$ by

\begin{align}
    K(t) =  \int_{0}^{t} k(s(t'))\frac{ds}{dt'}dt'. \label{eqn:total_curvature}
\end{align}

$K$ is related to the (signed) number of turns made by the curve via

\begin{equation}
    N_\mathrm{turns} = \frac{K}{2\pi}. \label{eqn:N_turns}
\end{equation}

Importantly, $K$ does \textit{not} measure the number of turns about the origin, or any fixed point in the $Q-U$ plane.  Rather, it computes the total amount of curvature in these curves from its intrinsic topology, invariant to translations and rotations of the curve.  We find that these properties are useful, as there are clear loops both in the data and simulations that do not revolve around, e.g., the mean values of $Q$ and $U$.  This also implies that our methodology is insensitive to the potential presence of an external Faraday screen.

For a light curve lasting from time $t_1$ to $t_2$, we estimate the average $Q-U$ rotation rate via

\begin{equation}
    \Omega_{QU} = \frac{K(t_2)-K(t_1)}{t_2-t_1}.
    \label{eqn:rotation_rate}
\end{equation}

We estimate both the first and second derivatives for \autoref{eqn:local_curvature} using three neighboring points.  The main challenges when applying this technique are (i) finite temporal sampling, for both the observational data and models, (ii) errors on observational data, and (iii) scrambled polarization for some Faraday-thick models.  For both the data and the simulations, we place a ``speed limit'' of $2\pi / 5 \ \mathrm{rad}\,\mathrm{t_g}^{-1}$.  That is, we zero the curvature whenever \autoref{eqn:rotation_rate} estimates that the curve locally makes at least one full rotation over the equivalent of one time-step in the simulations.  In practice, this affects less than 1\% of the data points in the observational data, and likewise for most of the simulations, but is important for catching numerical errors, since we are computing second derivatives.  

\autoref{fig:example} illustrates and summarizes our analysis pipeline.  We display the MAD, $a_\bullet=0.5$, $R_\mathrm{high}=1$, $i=30^\circ$ model between 21250 $t_g$ and 22250 $t_g$.  In the top row, log-scale images are presented with two decades in dynamic range.  Ticks are used to visualize the strength and EVPA of linear polarization.  Tick lengths scale with the total amount of polarization in a region ($\sqrt{Q^2+U^2}$), while tick colors scale with the fractional polarization ($|m|=\sqrt{Q^2+U^2}/I$).  On the bottom left, we plot the evolution of $Q$ and $U$ over this time period.  Colors are used to help map behavior to time in the bottom right plot, which displays $N_\mathrm{turns}$ (\autoref{eqn:N_turns}).

Between 21250 $t_g$ and 22750 $t_g$, the model is in a typical quiescent state without obvious distinct hotspots, yet we observe continuous looping behavior in the $Q-U$ plane.  At an inclination of $30^\circ$, the disk in this model moves counter-clockwise on the sky, and loops of the same handedness are found in $Q$ and $U$.  Two flux eruption events occur around 21750 $t_g$ and 22000 $t_g$, and as reported in \citet{Najafi-Ziyazi+2024} for some similar models, we notice significant polarization at the boundary of this expanding flux bubble.  During this period, the model makes a wider arc than usual in the $Q-U$ plane, eventually settling in a different area.  Plotting an ``X'' at the mean value of $Q$ and $U$ during this period, we find that the model makes zero loops around the mean during this time interval.  Despite this, \autoref{eqn:total_curvature} correctly captures many rotations that are evident by eye, as shown in the bottom right panel.  As illustrated by the dashed black line, the average slope is used to obtain $\Omega_{QU} = 4.6 \ \mathrm{deg} \, t_g^{-1}$ from this curve.

In this case, we have selected a relatively clean example whose loops are easy to follow by eye.  Consequently, $N_\mathrm{turns}$ increases nearly linearly with time.  An interesting exception occurs near the flux eruption event at 22000 $t_g$, where this disturbance in the accretion flow results in temporary clockwise motion in the $Q-U$ plane, resulting in a temporary decrease in $N_\mathrm{turns}$ before clockwise motion resumes.  As we will explore in upcoming sections, models with larger inclination and/or larger $R_\mathrm{high}$ depart more strongly from smooth linear evolution, which we attribute to ``noise'' from internal Faraday rotation.

\section{Light Curve Analysis}
\label{sec:light_curves}

\begin{figure*}
    \centering
    \includegraphics[width=\textwidth]{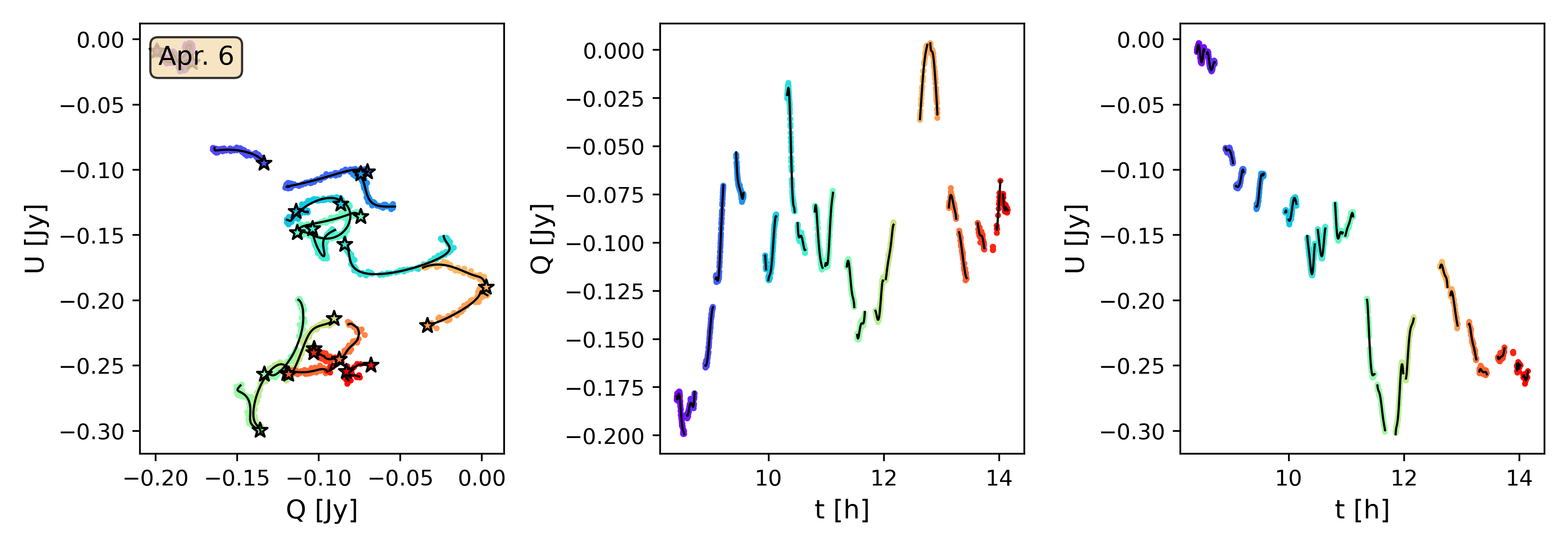}
    \includegraphics[width=\textwidth]{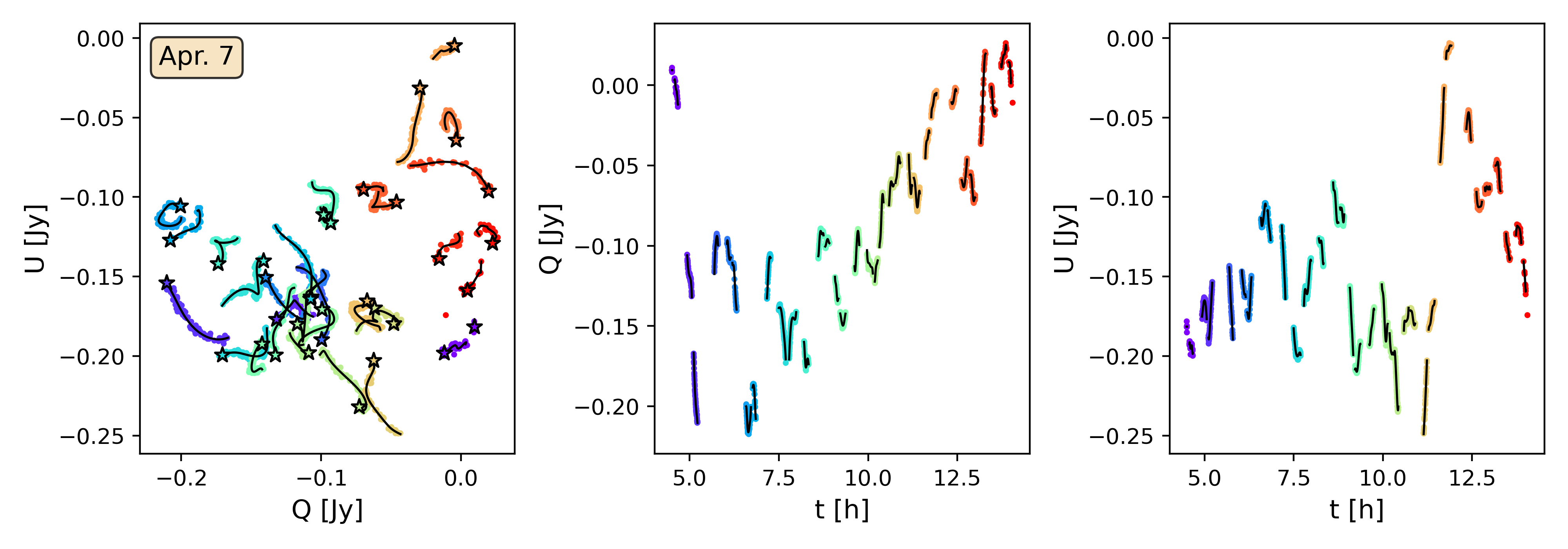}
    \includegraphics[width=\textwidth]{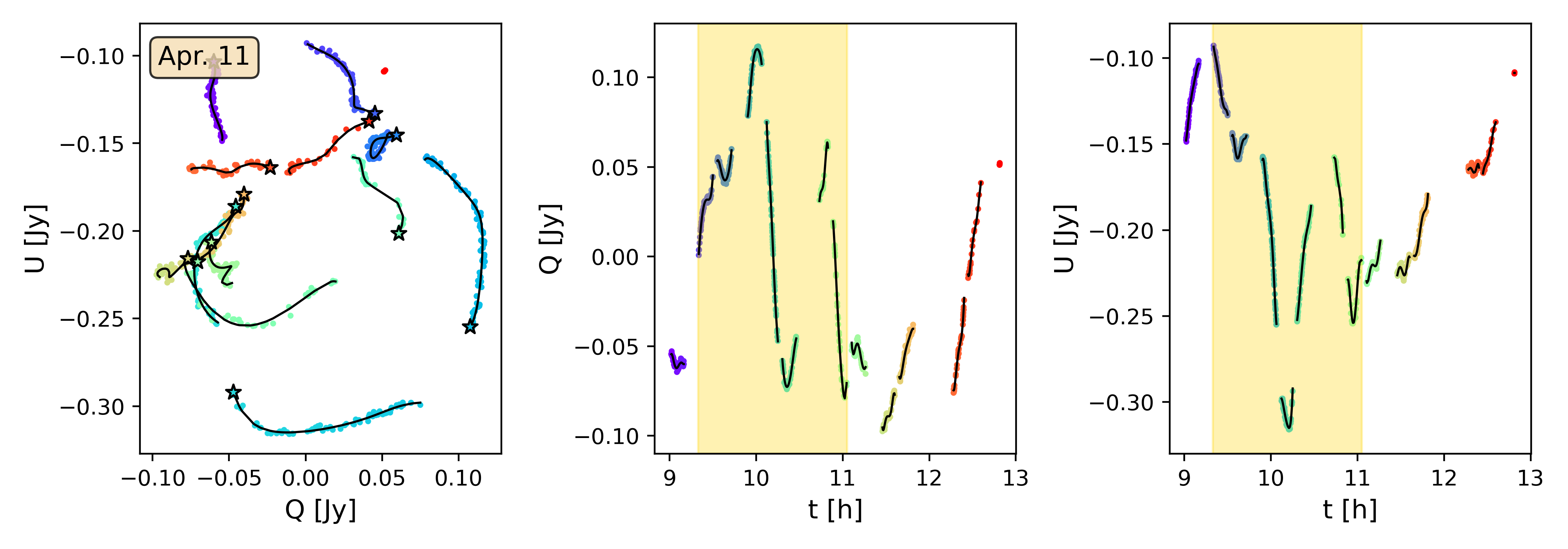}
     \caption{Sgr A* polarized light curves contemporaneous with the EHT 2017 observing campaign \citep{Wielgus2022QU}.  As illustrated here, we fit smoothing splines to each scan for curvature analysis.  We plot a star at the end of each spline to orient each curve.  The visually loopy period labeled $T_0$-$T_7$ in \citet{Wielgus2022QU} is highlighted in yellow.}
    \label{fig:light_curves}
\end{figure*}

We apply the the methodology outlined in \autoref{sec:methodology} to estimate the average rotation rate of the Sgr A* millimeter light curves presented in \citet{Wielgus2022QU} and visualized in \autoref{fig:light_curves}.  These data were obtained at 212.1-216.1 GHz and 226.1-230.1 GHz as part of the EHT+ALMA observations of \sgra \citep{EHTC+2022a}, and we focus on the highest frequency sub-band 228.1-230.1 GHz, which is the least affected by optical/Faraday depth.  The observational data are sampled with a nominal sampling rate of $4 \ \mathrm{s} = 0.2 \ t_g$, but with thermal noise and frequent gaps typically lasting a few minutes \citep{Wielgus+2022}.  Recall that first and second derivatives are both estimated using three adjacent data points, which could easily be disrupted by thermal noise. To smooth the data, we pre-process these light curves in the following manner:

\begin{enumerate}
    \item We split these light curves into segments, breaking them whenever there is a gap of at least 1.8 min to capture breaks between scans.
    \item To each segment of $Q(t)$ and $U(t)$, we fit a smoothing spline of polynomial order $\kappa \in \{3,4,5\}$ with smoothing factor $\sigma \in \{0.1,0.5\}$ using the Python function \texttt{scipy.interpolate.UnivariateSpline}.
    \item We replace each of the values of $Q(t)$ and $U(t)$ with their spline-interpolated values.
    \item Because high-order splines may acquire spurious curvature particularly on the edges of the fitting domain, we trim $n_T \in \{0,1,2,3\}$ data points from the ends of each of the segments.
    \item On each of the pre-processed scans, which last about 8 minutes on average, we apply equations \ref{eqn:arclength}-\ref{eqn:rotation_rate} to estimate the average $Q-U$ rotation rate.
\end{enumerate}

This methodology is unaffected by gaps in the data, so we need not make any assumptions regarding their continuity during gaps.  We visualize the light curves and the corresponding fitted smoothing splines (with $\kappa=5$, $\sigma=0.1$, and $n_T=0$) in \autoref{fig:light_curves}.  We find that these splines are generally able to characterize the $Q$ and $U$ behavior well, without obvious over-fitting.  We highlight in yellow the time period on April 11th extensively studied for its obvious looping behavior \citep[labeled as $T_0$ to $T_7$ in][]{Wielgus2022QU}.

This pre-processing procedure introduced several meta-parameters: $\kappa \in \{3,4,5\}$, $\sigma \in \{0.1,0.5\}$, and $n_T \in \{0,1,2,3\}$.  We found that our recovered $Q-U$ rotation rates are modestly sensitive to the values adopted.  Therefore, we survey each combination of parameters listed and estimate $\Omega_{QU}$ using those parameters.  We report the mean and use the standard deviation of these values as a systematic error bar in \autoref{tab:real_rotation_rates}, where we also provide the fraction of scans oriented clockwise from the sign of $\Omega_{QU}$, which we term $f_{CW}$.  

\begin{table}[h]
\centering
\begin{tabular}{|l|l|l|}
\hline
Time Interval        & $\Omega_{QU} \ [\mathrm{deg} \, t_g^{-1}]$ & $f_{CW}$ \\
\hline
April 6th          & $-2.8 \pm 0.4$ & $0.62 \pm 0.05$                                           \\
April 7th          & $-2.7 \pm 0.2$  & $0.65 \pm 0.05$                                        \\
April 11th & $-2.5 \pm 0.9$  & $0.69 \pm 0.13$                                          \\
April 11th ($T_0$-$T_7$) & $-4.0 \pm 0.7$ & $0.82 \pm 0.09$                                           \\
All Days           & $-2.6 \pm 0.6$  & $0.65 \pm 0.09$ \\
All Days (5$t_g$ pre-smoothed)       & $-2.8 \pm 1.0$  & $0.67 \pm 0.11$ \\
\hline
\end{tabular}
\caption{Polarization rotation rates and clockwise fraction of scans calculated from Sgr A* 2017 light curves.  Systematic error bars (1$\sigma$) are estimated by surveying meta-parameters associated with spline pre-processing of the data.  Although the visually loopy period $T_0$-$T_7$ is the most significantly biased towards clockwise, we find a bias towards clockwise motion during all time periods at a similar average speed.  Our ``All Days (5$t_g$ pre-smoothed)'' quantities are compared with GRMHD simulations.}
\label{tab:real_rotation_rates}
\end{table}

We consistently recover clockwise motion in the $Q-U$ plane, that is $\Omega_{QU} < 0$, for all time intervals of data, not only April 11th.  We obtain $\Omega_{QU} = -2.6 \pm 0.6 \ \mathrm{deg} \, t_g^{-1}$ and $f_{CW} = 0.65 \pm 0.09$ for our all-day fit, which corresponds to a period of 46 minutes.  $\Omega_{QU}$ is a timescale dependent quantity (see also \autoref{sec:app_timescales}), so to treat the data on the same timescale as the simulation cadence, we also compute an all-day fit on light curves first smoothed with a boxcar filter with duration 5$t_g$, resulting in $\Omega_{QU} = -2.8 \pm 1.0 \ \mathrm{deg} \, t_g^{-1}$ and $f_{CW} = 0.67 \pm 0.11$.  Thus, this additional smoothing did not significantly alter our recovered quantities.  

These data included multiple sub-band measurements between 213 and 229 GHz. While these data exhibit a rotation measure, the data between bands are highly correlated, and our methodology is constructed to be invariant to global translations and rotations in the $Q-U$ plane. In the 213 GHz sub-band, we similarly find clockwise motion on all days, with $\Omega_{QU} = -3.3 \pm 0.8 \ \mathrm{deg} \, t_g^{-1}$ and $f_{CW} = 0.68 \pm 0.09$ across all days, consistent within $1\sigma$.

For comparison, a joint hotspot fit of six NIR flares yielded an angular velocity of approximately $-6 \ \mathrm{deg}\,\mathrm{min}^{-1}$ \citep[][]{Gravity+2023}. Accounting for the fact that two $Q-U$ loops occur during one revolution in their model (even if, as they argue, one loop is potentially too small to be caught by their observing cadence), this equates to a $Q-U$ loop speed of approximately $-4 \ \mathrm{deg}\,\mathrm{t_g}^{-1}$ (recalling that $1 \ \mathrm{min} \approx 3 \ t_g$ for \sgra).  This is larger than our all-day fit value, but agrees exactly with the loop speed found during the post-flare period. On April 11th, we find a characteristic period of 48 minutes, or 30 minutes when considering only the ``loopy'' period between $T_0$ and $T_7$.  During this time period of 103 minutes, \citet{Wielgus2022QU} report two large loops and one small one:  three loops within 103 minutes is consistent with the rotation speeds calculated here.  It is during this period that the clockwise fraction of scans is highest, $f_{CW} = 0.82 \pm 0.09$.

\section{Model Comparison}
\label{sec:model_comparison}

\begin{figure*}
    \centering
    \includegraphics[width=\textwidth]{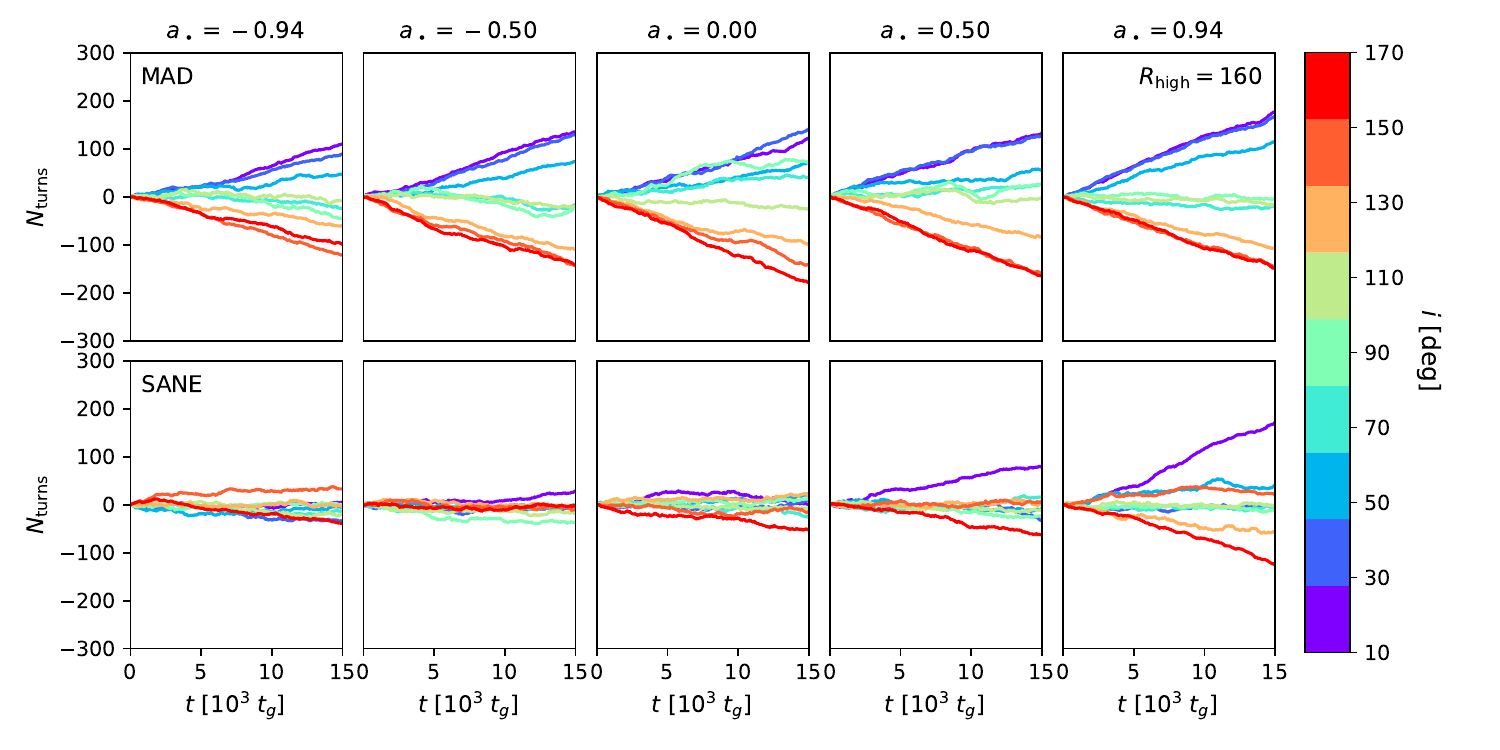}
    \caption{$N_\mathrm{turns}$ as a function of time for the models in our GRMHD library with fixed $R_\mathrm{high}=160$.  Although these curves are non-monotonic, we find clear slopes whose signs encode the line-of-sight inclination.  For many SANE models with such a large value of $R_\mathrm{high}$, the linear polarization is sufficiently scrambled by Faraday rotation to erase a signal that may otherwise be evident at smaller $R_\mathrm{high}$ values (see \autoref{fig:theory_comparison_omega}).}
    \label{fig:examples_fixedRhigh}
\end{figure*}

Because our GRMHD simulations have no thermal noise and no data gaps, we do not perform spline pre-processing and apply equations \ref{eqn:arclength}-\ref{eqn:rotation_rate} directly to their simulated light curves.  In \autoref{fig:examples_fixedRhigh}, we first plot $N_\mathrm{turns}(t)$ for all of the models in our library with fixed $R_\mathrm{high}=160$.  These curves are non-monotonic, but they exhibit a clear long-term slope from which $\Omega_{QU}$ can be estimated.  The clearest trend is with respect to inclination, with negative slopes if $i < 90^\circ$ and positive slopes if $i>90^\circ$.  The handedness of the accretion flow can therefore directly be inferred from the sign of $\Omega_{QU}$, where clockwise motion in the $Q-U$ plane is associated with clockwise motion in the accretion flow on the sky.  It is non-trivial that it is the accretion disk's inclination rather than the BH spin that governs this pattern: this is consistent with the behavior observed for Stokes I pattern speeds from these simulations \citep{Conroy+2023}.  Hotspot models provide insight into the smooth evolution with respect to inclination.  Face-on hotspots usually produce two $Q-U$ loops (one big one and one small one) per orbit.  As the inclination increases, the smaller one may shrink or disappear, depending on the details of the orbit and magnetic field geometry \citep[e.g.,][]{Gravity2018QU,Gelles+2021,Vincent2024}.

For $R_\mathrm{high}=160$, the difference between MAD and SANE models is mainly due to Faraday depolarization, not necessarily due to differences in the underlying dynamics.  SANE models are typically much more Faraday thick (due to higher densities and lower temperatures), leading to smaller polarization fractions as well as temporal decoherence \citep{EHTC+2024c}.  However, we notice that the more face-on SANE $a_\bullet=0.5$ and $a_\bullet=0.94$ images exhibit slopes similar to MADs.  This is because these images exhibit more jet emission than their lower-spin counterparts.  If these models are face-on ($i=10^\circ$ or $i=170^\circ$), emission from the forward-jet can reach the observer without passing through a Faraday-thick region, allowing for transmission of this polarized signature.  Models with smaller $R_\mathrm{high}$ are less Faraday thick, and this signal can therefore be more accessible for such models.

\begin{figure*}
    \centering
    \includegraphics[width=\textwidth]{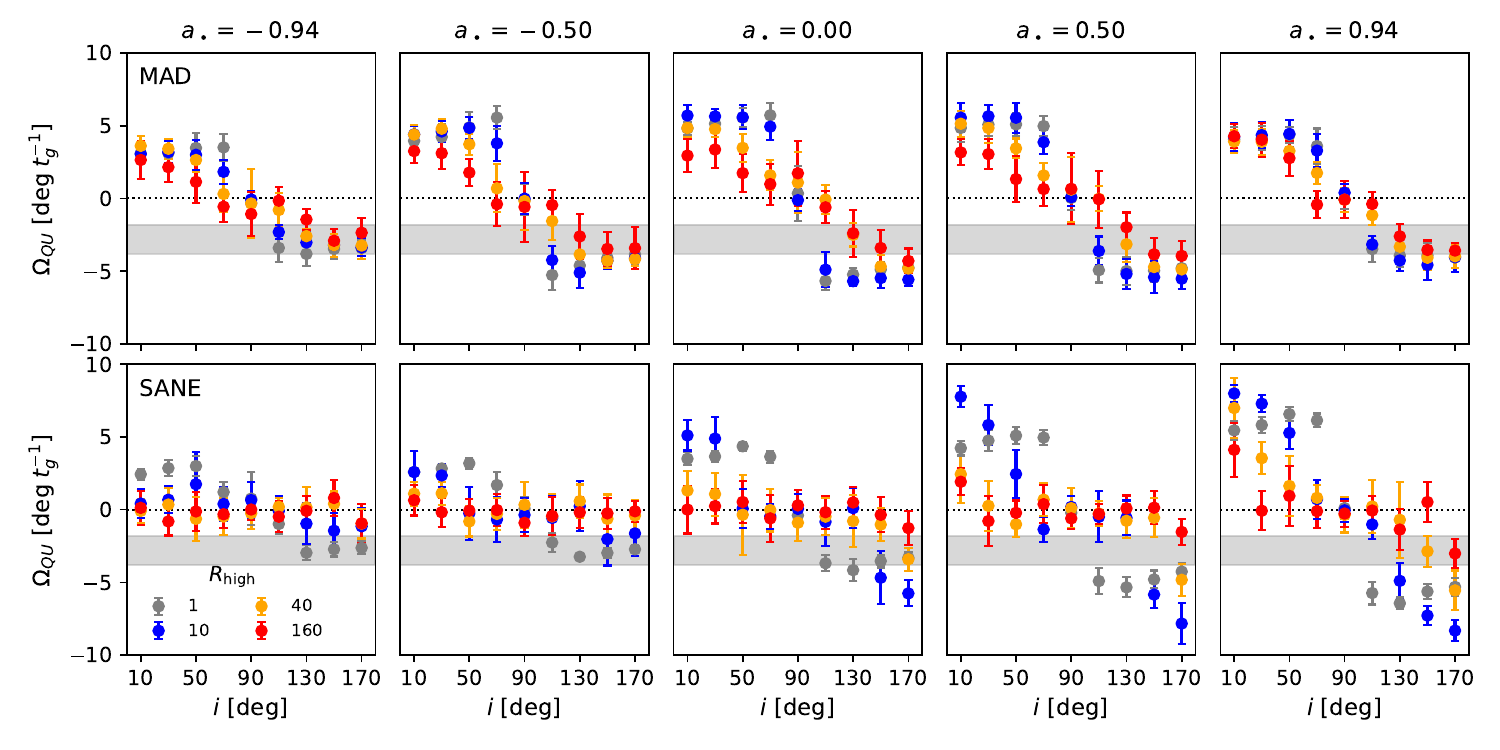}
    \caption{Polarization rotation rates estimated from our GRMHD simulations.  Error bars represent the standard deviation obtained among 10 windows of 1500 M.  In grey, we plot the ``All Days (5$t_g$ pre-smoothed)'' constraint from \autoref{tab:real_rotation_rates}.  Clockwise MAD models are most likely to pass this constraint, including the ``best bet'' model identified in \citet{EHTC+2022e,EHTC+2024c}.}
    \label{fig:theory_comparison_omega}
\end{figure*}

\begin{figure*}
    \centering
    \includegraphics[width=\textwidth]{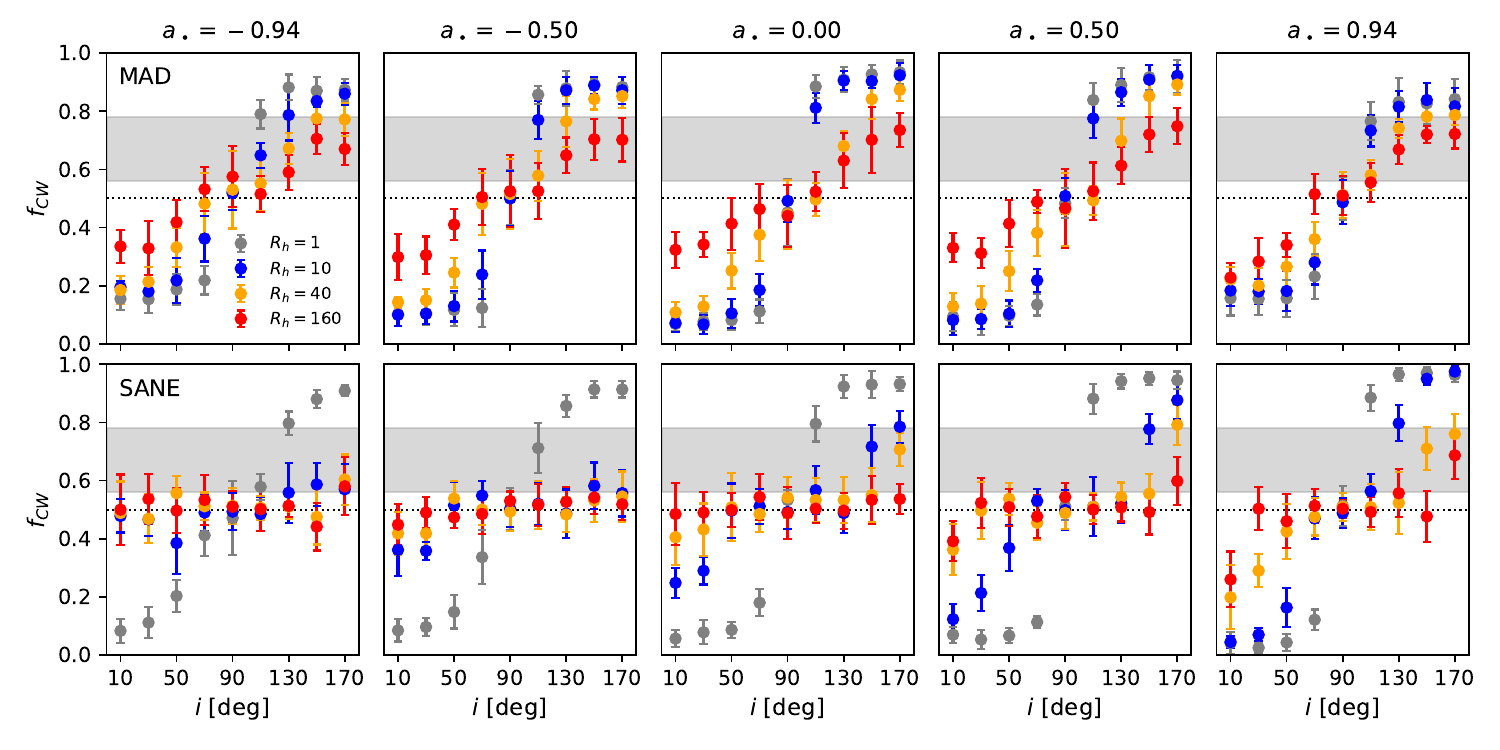}
    \caption{As \autoref{fig:theory_comparison_omega}, but for the fraction of time that segments of length $25 \ t_g$ are clockwise, compared with observational data.}
    \label{fig:theory_comparison_clockwise}
\end{figure*}

Because $N_\mathrm{turns}(t)$ is non-monotonic, and the observational data are sampled across three days, we estimate $\Omega_{QU}$ within 10 evenly-spaced windows across the 15,000 M simulated.  This is consistent with the methodology of \citet{EHTC+2024c}.  Each window is 8.3 hours in length, similar to the duration of one day's light curve.  From these 10 values of $\Omega_{QU}$, we plot the mean and standard deviation in \autoref{fig:theory_comparison_omega}.  Similarly, we consider the fraction of $25 \ t_g$ segments within each $1500 \ t_g$ window which are clockwise.  This potentially breaks a degeneracy between (i) slower motion in the $Q-U$ plane overall and (ii) a less pronounced bias towards either clockwise or counter-clockwise motion, each of which could lower $\Omega_{QU}$. Results from this calculation are provided in \autoref{fig:theory_comparison_clockwise}. In both figures, we over-plot the $1\sigma$ ``All Days (5$t_g$ pre-smoothed)'' measurement from \autoref{tab:real_rotation_rates} in grey.

From this model comparison, we find that face-on and clockwise ($i>90^\circ$) MAD models are most likely to match observational constraints.  SANE models, generally disfavored by EHT studies \citep{EHTC+2022e,EHTC+2024c}, are typically too depolarized to produce large enough rotation rates to match the data.  Edge-on models are more likely to be either depolarized or produce equal amounts of clockwise and counter-clockwise loops.  The ``best bet'' model of \citet{EHTC+2024c}, MAD, $a_\bullet=0.94$, $R_\mathrm{high}=160$, $i=150^\circ$, with an aligned magnetic field, lies within the $1\sigma$ observational measurement, in excellent agreement.

\begin{figure*}
    \centering
    \includegraphics[width=\textwidth]{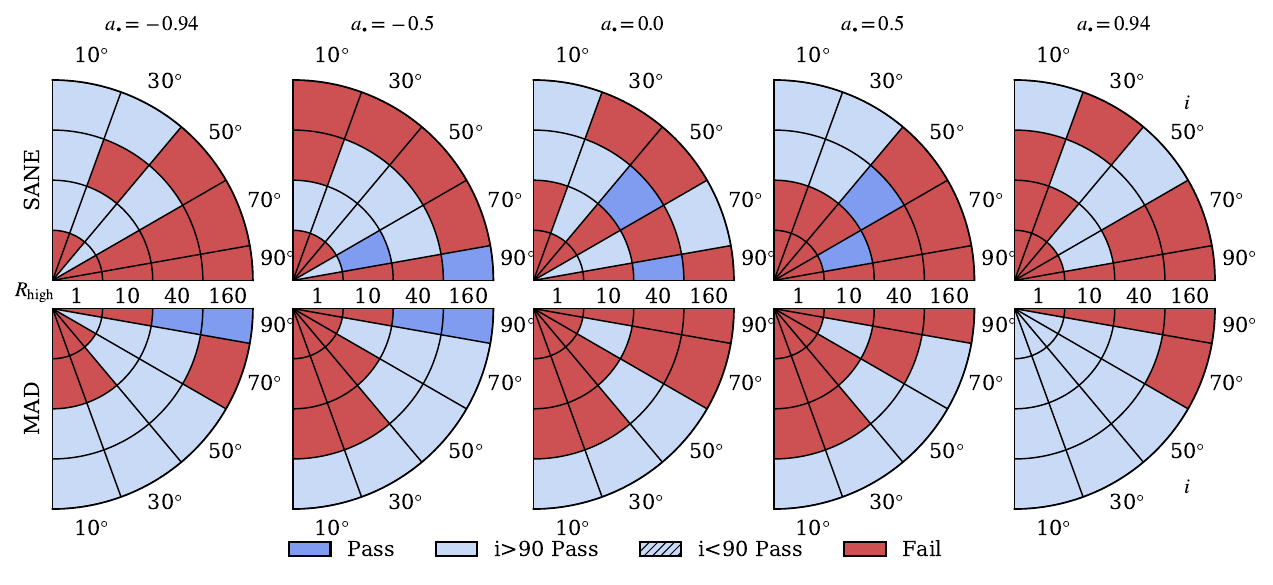} 
    \caption{Pizza plots visualizing our constraints on parameter space, based on both $\Omega_{QU}$ and the $f_{CW}$. Models are marked as passing if the 1$\sigma$ error bars of the model fall within the 1$\sigma$ data regions in both \autoref{fig:theory_comparison_omega} and \autoref{fig:theory_comparison_clockwise}.  Clockwise models ($i>90^\circ$) are much more likely to pass than their counter-clockwise ($i<90^\circ$) counterparts.  The ``best bet'' model of \citet{EHTC+2024c} passes our new constraints.}
    \label{fig:pizza}
\end{figure*}

In \autoref{fig:pizza}, we visualize a pass/fail table based on the measured $Q-U$ rotation rate and clockwise fraction.  Models pass if their 1$\sigma$ confidence region overlaps with the 1$\sigma$ confidence region of observations.  Here, clockwise $i>90^\circ$ models are much more likely to pass observational constraints than their $i<90^\circ$ counterparts.  This trend is clearer for MADs than for SANEs, where several models exhibit larger standard deviations with means closer to zero.  As implied by the structure of the curves shown in \autoref{fig:examples_fixedRhigh}, if longer time periods are considered, smaller theoretical errors are obtained. This implies that it would be valuable to continue to accrue longer light curves of \sgra, as well as produce longer light curves from GRMHD simulations.  

\section{Discussion}
\label{sec:discussion}

\subsection{Inconclusive Results at 86-100 GHz}

On April 3rd, 2017, polarized light curves in 85.3-89.3 GHz and 97.3-101.3 GHz were also recorded as a part of GMVA+ALMA observations of \sgra \citep{Issaoun2019}. The light curves were reduced and published in \citet{Wielgus+2024}. In \autoref{fig:86}, we plot and quote values for the highest frequency sub-band at 100.3 GHz, which should be the least affected by Faraday effects.  By eye, the data appear to exhibit a global counter-clockwise loop.  However, such global loops are not necessarily detected by our method, since we integrate only the local curvature without assuming continuity between scans.

Repeating the analysis of \autoref{sec:light_curves} on these data, we obtain inconsistent sub-band dependent results:  $\Omega_{QU} = -0.1 \pm 1.2 \ \mathrm{deg}\,\mathrm{t_g}^{-1}$ at 86 GHz, $\Omega_{QU} = 0.4 \pm 1.0 \ \mathrm{deg}\,\mathrm{t_g}^{-1}$ at 88 GHz, $\Omega_{QU} = 0.9 \pm 1.0 \ \mathrm{deg}\,\mathrm{t_g}^{-1}$ at 98 GHz, and $\Omega_{QU} = 1.4 \pm 0.7 \ \mathrm{deg}\,\mathrm{t_g}^{-1}$ at 100 GHz.  That is, almost all sub-bands are consistent with 0, and only the highest frequency sub-band is $2\sigma$ inconsistent with 0, with counter-clockwise motion obtained.  We therefore report inconclusive results at this band.

Compared to 230 GHz, we expect the innermost regions to be more obscured due to increased optical depth, and larger Faraday rotation at this frequency may further corrupt the signal.  There is also less usable data at 86-100 GHz than at 230 GHz; continued monitoring may allow us to uncover hints of dynamics in longer datasets.  Our spline fits appear to exhibit more substructure compared to 230 GHz, and this could mean that there are too many small wiggles and gaps in the data to successfully extract the $Q-U$ loop speed by integrating the local curvature.  This motivates a refinement of our treatment of data gaps and noise in future work.

\subsection{Other Dynamical Tracers}

\begin{figure*}
    \centering
    \includegraphics[width=\textwidth]{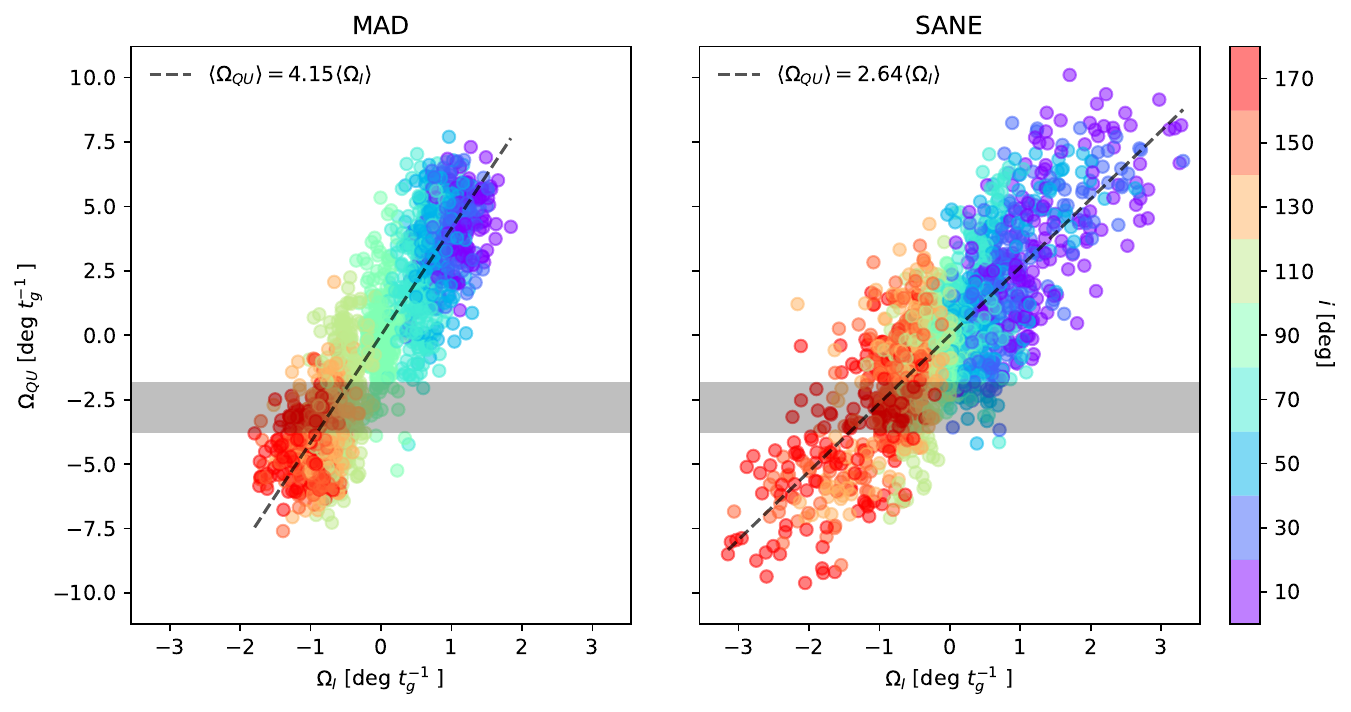}
    \caption{Comparison of polarization loop speed ($\Omega_{QU}$; this work) and pattern speed in Stokes I \citep[$\Omega_{I}$;][]{Conroy+2023}.  We find linear correlations, but separate ones for MADs and SANEs. The shaded region corresponds to the range within a standard deviation of the observational value (all days, $5t_g$ pre-smoothed).}
    \label{fig:pattern_speed}
\end{figure*}

\citet{Conroy+2023} studied the Stokes $I$ rotational pattern speeds of the same simulated images used in this study, allowing a direct comparison with the $Q-U$ loop speeds calculated here.  They found that the pattern speed was more sensitive to inclination than to spin, and that MAD and SANE simulations followed different relationships, as in our study.  In \autoref{fig:pattern_speed}, we plot a comparison of the $Q-U$ loop speed from our study and the pattern speed computed as in \citet{Conroy+2023} for each of the 1500 $t_g$ time intervals computed in \autoref{sec:model_comparison}. 

Since both tracers depend on inclination, both tend to rotate in the same direction. Models that pass the $Q-U$ loop speed and clockwise $Q-U$ fraction constraints in \autoref{fig:pizza} also generally exhibit a clockwise pattern speed. Passing MAD models have a pattern speed of $\overline{\Omega}_I = -0.6\pm 0.4$ $\mathrm{deg} \, t_g^{-1}$, while passing SANEs have a pattern speed of $\overline{\Omega}_I = -0.4\pm 0.6$ $\mathrm{deg} \, t_g^{-1}$. Thus, we would predict a strongly sub-Keplerian clockwise pattern speed based on measurements of $\Omega_{QU}$. Our prediction is consistent with preliminary dynamical reconstructions of Sgr A* on April 6 2017, which exhibit a pattern speed of $\Omega_I\approx-0.7$ $\mathrm{deg} \, t_g^{-1}$ \citep{Knollmueller+2023}. Future EHT analysis, with full validation across multiple imaging methods, will enable robust tests of this prediction.

We find a remarkably linear trend between $\Omega_{QU}$ and $\Omega_I$, considering that Faraday rotation offers a mechanism to affect the $Q-U$ loop speed without changing the Stokes $I$ pattern speed.  MAD models follow a steeper relationship than their SANE counterparts.  Both slopes are larger than the most natural value, 2, since the hotspot will typically make two $Q-U$ loops \citep[one large and one small][]{Gelles+2021,Vos2022,Vincent2024} during one orbit for the rotationally symmetric magnetic field configurations consistent with the observations \citep[e.g.,][]{Gravity2018QU,Wielgus2022QU,Gravity+2023}.  

\begin{figure}
    \centering
    \includegraphics[width=0.45\textwidth]{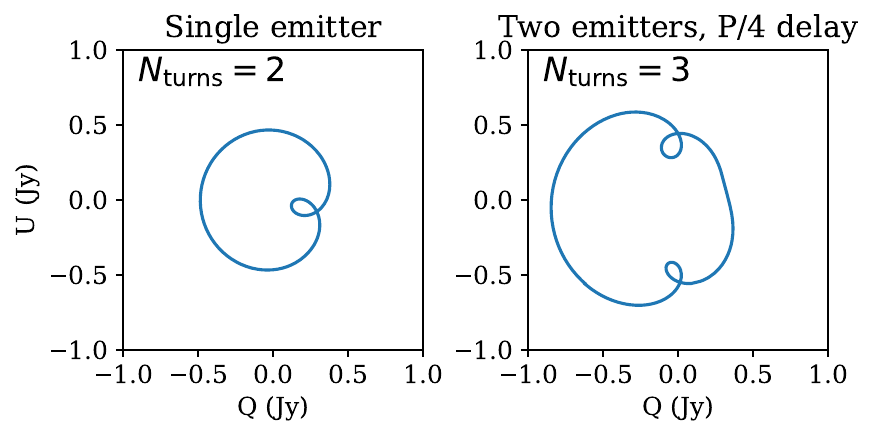}
    \caption{{\it Left}: $Q-U$ loop from a single hotspot moving on a circular orbit around a Schwarzschild black hole at the innermost stable circular orbit with a vertical magnetic field. {\it Right}: same as left, but a second identical emitter is orbiting simultaneously at a quarter period delay. Though the detailed features of these curves depend on the model, additional emitters can generally change curvature and add turns to the observed loop while preserving the pattern speed.}
    \label{fig:timedelay}
\end{figure}

\begin{figure*}
    \centering
    \includegraphics[width=\textwidth]{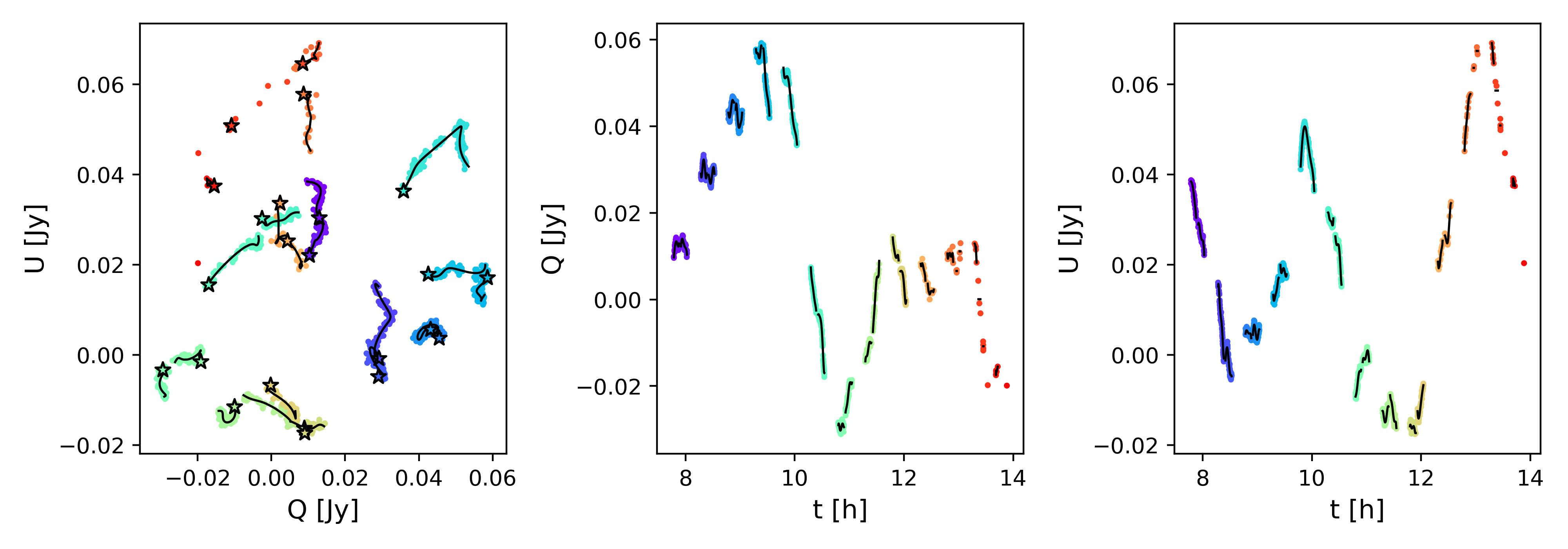}
    \caption{As \autoref{fig:light_curves}, but for 100 GHz light curves taken during April 3rd, 2017 \citep{Issaoun2019,Wielgus+2024}.  We obtain hints of slower clockwise motion, but $\Omega_{QU}$ is consistent with 0 within our systematic error bars.}
    \label{fig:86}
\end{figure*}

One explanation may be that multiple polarized perturbations with the same pattern speed can each produce independent loops.  In \autoref{fig:timedelay}, we present an example of this scenario generated using the code \texttt{KerrBAM} \citep{Palumbo_2022}.  Here, we model a hotspot on a closed orbit around a Schwarzschild black hole at a radius of 6 $r_g$, with a speed of $0.4c$ in the Boyer-Lindquist zero angular momentum observer frame. The magnetic field is vertical (out of the mid-plane) in the fluid frame, while the mid-plane normal is oriented $30^\circ$ from the observer line of sight. The emitter itself is an optically thin rigidly rotating Gaussian blob with a full width at half maximum of $1 r_g$ for which the total flux and polarization fraction are both arbitrarily selected, as \texttt{KerrBAM} does not model radiative transfer. Geodesics are terminated at the first mid-plane crossing to avoid strong lensing effects from secondary images in this particular test.

In the $Q-U$ plane, we add the signal from hotspot to a copy of itself with the same flux density that is time-delayed by a quarter of the period.  This represents a second perturbation in the accretion flow on the same orbit, but out of phase, which would therefore preserve the same pattern speed.  We see that this results in additional loop in the $Q-U$ plane, increasing the $Q-U$ loop speed by 50\%.  In general, less idealized setups (including a radial dependence in the velocity/magnetic field sampled by the emission) could naturally form even more loops in the $Q-U$ space.  Consequently, it is natural to expect a slope larger than 2, but as a result it is not straightforward to directly map $\Omega_{QU}$ to a dynamical feature of the accretion flow without assuming restrictions on the emission geometry.

In addition to the polarization loop speed ($\Omega_{QU}$, this work) and pattern speed in Stokes I \citep[$\Omega_I$, see][]{Conroy+2023}, others have proposed methods to detect near-horizon dynamics using correlated visibility amplitudes across tracks \citep[$\Omega_{uv}$, see Conroy et al. in prep., ][]{Johnson+2015}, centroid motion \citep{Gravity2018QU}, and even fluctuations in the polarization fraction along a baseline \citep{Fish+2009}. While we might detect near-horizon dynamics using any of these methods, our study highlights that these methods may be sensitive to different features and scales in the system. As exemplified by \autoref{fig:pattern_speed}, if the accretion disk behaves as expected from typical GRMHD (e.g. numerous simultaneous fluctuations, non-stationary radial velocity profiles and magnetic field profiles, etc.), then we might not expect the inferred rotation rates to relate by a predictable analytic factor {\it a priori}. As a corollary, if measurements of $\Omega_{QU}$ do not differ by a simple factor of $2\times$ from measurements of $\Omega_I$ or $\Omega_{uv}$, then we might infer the emission structure is behaving as expected from typical GRMHD. Alternatively, if these measures do differ by a simple factor of $2\times$, then we might infer that the emission can be explained by a single, large-amplitude fluctuation.  

\subsection{Indirect Support for an External Faraday Screen}

Although our methodology is insensitive to an external Faraday screen, the clockwise motion that we recover provides supporting evidence for such a screen in the context of other studies.  The ``best bet'' model for \sgra identified in \citet{EHTC+2022e,EHTC+2024c} is oriented at $i=150^\circ$ despite the fact that the handedness of the polarization pattern ($\angle \beta_2$) naturally implies counter-clockwise motion.  \citet{EHTC+2024c} argued that this discrepancy could be due to an external Faraday screen.  Clockwise motion inferred from both the $Q-U$ plane and centroid motion in the NIR provided additional circumstantial evidence for the external Faraday screen interpretation, but these were based on hotspot models during flaring periods \citep{Gravity2018QU,Wielgus2022QU,Gravity+2023}.  One of the key concerns with this interpretation had been evidence of {\it internal} Faraday rotation from rotation measure (RM) variability timescales \citep{Wielgus2022QU}, 86 GHz measurements \citep{Wielgus+2024}, and the simulations themselves \citep{Moscibrodzka+2017,Jimenez-Rosales&Dexter2018,Ricarte+2020,Ressler+2023}.  Our analysis of the 2017 \sgra light curves now provides evidence for clockwise rotation accretion flow during the exact same non-flaring days which were imaged, further supporting this interpretation.  \citet{Antonopoulou+2025} presented a model in which a counter-clockwise rotating disk produced clockwise hotspots traveling along magnetic flux tubes, which would have opposite helicity.  Our analysis appears to disfavor this model, which we predict would produce counter-clockwise $Q-U$ loops on non-flaring days.

\subsection{Other Historical $Q-U$ Loops}

Because measuring $\Omega_{QU}$ provides an inclination constraint on \sgra, $Q-U$ loops in different epochs offer insights into the stability of the accretion flow, which could potentially have implications for the BH fueling mechanism and spin evolution \citep[e.g.,][]{Berti&Volonteri2008,Wang&Zhang2024,Ricarte+2025}. \citet{Gravity2018QU} and \citet{Gravity+2023} reported clockwise motion from NIR $Q-U$ loops and centroid motion between 2018 and 2022, suggesting that the orientation of the accretion flow with respect to our line-of-sight has remained stable on a timescale of $6 \ \mathrm{yr} \sim10^7 \ t_g$.  The consistently negative sign of circular polarization over decades provides additional indirect evidence of structural consistency in the accretion flow \citep{Munoz+2012,Bower+2018,Wielgus2022QU,Michail+2023}, although its interpretation is complicated by unknown contributions from Faraday conversion and intrinsic emission \citep{Ricarte+2021b,EHTC+2023,EHTC+2024c,Joshi+2024}.  This may already provide a valuable constraint on GRMHD simulations that do not assume an aligned or anti-aligned accretion disk in the initial conditions \citep[e.g.,][]{Chatterjee+2020,Ressler+2020,Olivares+2023}.  Intriguingly, \citet{Marrone2006} reported a {\it counter}-clockwise $Q-U$ loop in 2005 over the course of 3.5 hours, albeit with an observational cadence of tens of minutes.  While this is suggestive of a flip of the accretion flow geometry on a timescale of $\sim$$2\times 10^7 \ t_g$, we caution that occasional counter-clockwise loops are expected even in our simulations that assume perfect alignment or anti-alignment.  Therefore, continued high-cadence monitoring of \sgra will be crucial for providing constraints on the accretion flow geometry on a variety of time-scales.

\subsection{Handedness Flips}

\begin{figure*}
    \centering
    \includegraphics[width=\textwidth]{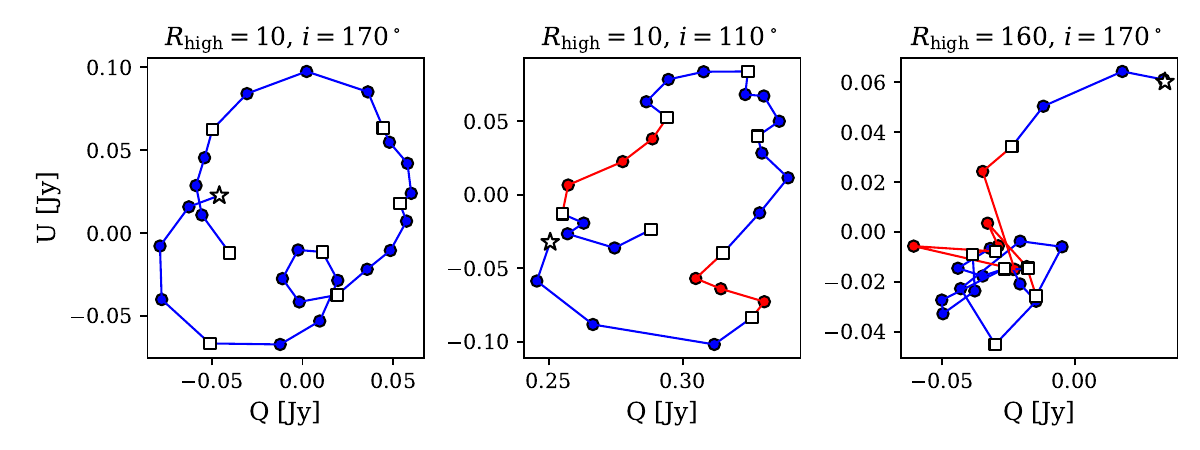} 
    \caption{$Q-U$ evolution of our MAD $a_\bullet=0$ simulations with different parameters, during an illustrative time interval of length $150 \ t_g$.  Segments of $25 \ t_g$ are separated by squares, with a star marking the latest time.  Each $25 \ t_g$ segment is colored either blue if clockwise curvature is calculated, or red if counter-clockwise curvature is calculated.  Compared to the smooth evolution in the $R_\mathrm{high}=10$, $i=170^\circ$ model, noisier substructure occurs at either $i=110^\circ$ or $R_\mathrm{high}=160$, occasionally flipping the curvature to clockwise.}
    \label{fig:handedness_example}
\end{figure*}

In \autoref{sec:model_comparison}, we found that the clockwise duty cycle carries similar information to the $Q-U$ rotation rate, but not identical.  For example, SANE $R_\mathrm{high}=1$ models with $i>110^\circ$ can pass the $\Omega_{QU}$ constraint, but usually have a $f_{CW}$ that is too high.  These models have large emission radii and are Faraday thin, making consistently clockwise but slower-than-average loops.  Among MAD models, the $f_{CW}$ constraint results in a mild preference for models with larger values of $R_\mathrm{high}$.  In such models, greater internal Faraday rotation can scramble what would otherwise be overly consistent clockwise loops.

In \autoref{fig:handedness_example}, to explore the underlying causes of these counter-rotating periods, we plot the $Q-U$ evolution of three of our MAD $a_\bullet=0$ simulations during an illustrative time interval of length $150 \ t_g$.  These models are selected to have decreasing $f_{CW}$ and correspond to $R_\mathrm{high}=10$ and $i=170^\circ$, $R_\mathrm{high}=10$ and $i=110^\circ$, and $R_\mathrm{high}=160$ and $i=170^\circ$.  Segments of length $25 \ t_g$ are demarcated with squares, and these curves are color-coded according to the curvature calculated within that segment: blue for clockwise, and red for counter-clockwise.  Note that because $R_\mathrm{high}$ and $i$ are specified in post-processing, these correspond to the exact same GRMHD fluid snapshots.

The $R_\mathrm{high}=10$ $i=170^\circ$ model makes clear, consistently clockwise loops that are easily followed by eye.  We find that this behavior is typical of face-on models with low $R_\mathrm{high}$.  The more edge-on $i=110^\circ$ case exhibits more substructure, likely due in part to increased Faraday rotation in the disk, but also in part to the unusual viewing angle causing loops to be partly concave, which transiently produces local curvature with the opposite handedness even for idealized hotspot models \citep[see e.g., Figure 8 of][]{Gelles+2021}.  The $R_\mathrm{high}=160$ $i=170^\circ$ model has significantly larger Faraday rotation depth than its $R_\mathrm{high}=10$ counterpart \citep{EHTC+2024c}.  Internal Faraday rotation produces much more noise-like structure, that occasionally produces curvature of the opposite sign.

Since all panels are the same GRMHD snapshots, it is clear that occasional periods of counter-clockwise motion do not correspond to global changes to the accretion flow.  Instead, turbulent motion, internal Faraday rotation, and light bending effects associated with less face-on viewing angles can naturally explain a non-unity $f_{CW}$.  We remark that $f_{CW}$ is timescale dependent, and as one would expect from \autoref{fig:examples_fixedRhigh}, it grows closer to unity (for clockwise models) as the timescale increases.

\section{Conclusion}
\label{sec:conclusion}

We have developed a method to characterize rotation rates of polarized light curves in the $Q-U$ plane, allowing measurement of a characteristic rotation speed and handedness even in regions that do not necessarily appear ``loopy'' by eye.  We have applied this technique to light curves of \sgra contemporaneous with the 2017 EHT campaign and to a library of GRMHD simulations for model comparison.  Our main results are as follows:

\begin{itemize}
    \item We measure clockwise motion in the $Q-U$ plane for all three days of observation in April 2017, not only on April 11th, where clockwise motion had previously been identified by eye.  We report that $65 \pm 9 \%$ of scans were biased clockwise, and an average $Q-U$ rotation rate of $-2.6 \pm 0.6 \ \mathrm{deg}\,\mathrm{t_g}^{-1}$.
    \item We find that $Q-U$ looping behavior is ubiquitous in our GRMHD library, with clockwise motion in the $Q-U$ plane corresponding to clockwise motion of features on the sky.  Notably, the handedness of loops in the $Q-U$ plane follows the inclination of the accretion disk, not the BH spin.  This is consistent with the behavior seen in Stokes $I$ pattern speeds \citep{Conroy+2023}.
    \item We use the $Q-U$ rotation rate and the clockwise duty cycle as constraints on our GRMHD library.  Face-on MAD models are most likly to pass these constraints, including the ``best-bet'' model identified in EHT theory studies \citep{EHTC+2022e,EHTC+2024c}.  This supports the interpretation in \citet{EHTC+2024c} of a Faraday screen along our line-of-sight that flips the handedness of the polarization pattern.  Although clockwise motion from previous $Q-U$ loops had already supported this interpretation, we strengthen this argument by measuring clockwise motion on the exact same days and observing frequencies as the EHT campaign. 
\end{itemize}

In principle, this technique could be applied to any BH with an optically thin hot accretion flow, with the relevant timescales scaling linearly with the mass.  This would result in model-dependent joint constraint on mass and inclination, where mass could be independently constrained via Stokes $I$ variability timescales \citep{Bower+2015,Chen+2023} or other BH mass estimators.  Although we did not explore higher frequencies in this work, based on \sgra's behavior at 86-100 GHz, we speculate that 345 GHz may present cleaner looping behavior, since both the optical and Faraday rotation depths would significantly decrease.  These findings motivate continued multi-wavelength high-cadence monitoring of \sgra to constrain the persistence of its accretion geometry.

\section{Acknowledgments}
We thank Roy Herrera for early explorations into looping behavior during flares in GRMHD simulations.  We thank Michi Baubock and Charles Gammie for sharing their insights into looping behavior.  We thank Sara Issaoun and Avery Broderick for useful discussions.  

This project/publication is funded in part by the Gordon and Betty Moore Foundation (Grant \#8273.01). It was also made possible through the support of a grant from the John Templeton Foundation (Grant \#62286).  The opinions expressed in this publication are those of the author(s) and do not necessarily reflect the views of these Foundations. NC is supported by the NASA Future Investigators in NASA Earth and Space Science and Technology (FINESST) program. This material is based upon work supported by the National Aeronautics and Space Administration under Grant No. 80NSSC24K1475 issued through the Science Mission Directorate. MW is supported by a Ramón y Cajal grant RYC2023-042988-I from the Spanish Ministry of Science and Innovation.

This paper makes use of the following ALMA data: ADS/JAO.ALMA\#2016.1.01404.V and ADS/JAO.ALMA\#2016.1.00413.V ALMA is a partnership of ESO (representing its member states), NSF (USA) and NINS (Japan), together with NRC (Canada), NSC and ASIAA (Taiwan), and KASI (Republic of Korea), in cooperation with the Republic of Chile. The Joint ALMA Observatory is operated by ESO, AUI/NRAO and NAOJ. 

\appendix

\section{Polarization Rotation Rates by Timescale}
\label{sec:app_timescales}

\begin{figure*}
    \centering
    \includegraphics[width=\textwidth]{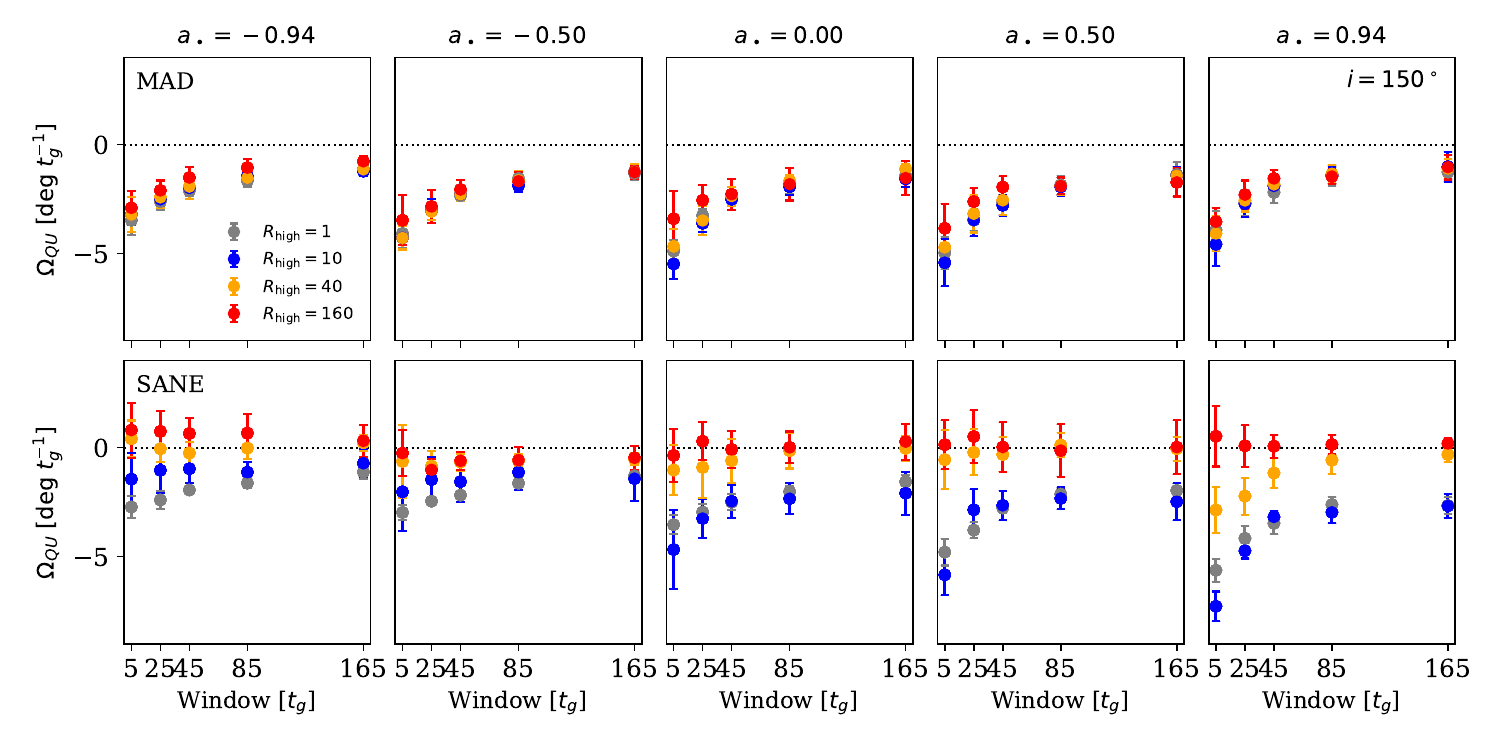}
    \caption{$\Omega_{QU}$ as a function of time for the models in our GRMHD library with $i=150^\circ$, if curves are pre-smoothed by a boxcar filter whose duration is noted on the x-axis.}
    \label{fig:smoothing_test}
\end{figure*}

Both GRMHD simulations and idealized hotspot models are capable of producing arbitrarily small and short-duration loops.  Because our models are ray-traced at a finite cadence of 5 $t_g$, this motivated consistent smoothing of the data with a 5 $t_g$ boxcar filter during theoretical comparison.  Here, we experiment with smoothing the GRMHD light curves with boxcar filters of different durations to investigate the structure of this evolution.  In \autoref{fig:smoothing_test}, we plot $\Omega_{QU}$ for our $i=150^\circ$ models if their Stokes $Q$ and $U$ curves are pre-smoothed with boxcar filters of duration 5 (no smoothing), 25, 45, 85, and 165 $t_g$.  For all of our MAD models, we find that increasing this window size decreases the magnitude of $\Omega_{QU}$, but $\Omega_{QU}$ never reaches 0 for the window sizes investigated.  SANE models exhibit overall similar behavior, but some models are consistent with $\Omega_{QU}=0$ for all window sizes investigated.

This highlights the timescale-dependence of $\Omega_{QU}$ and suggests that the models would exhibit additional loops that are temporally unresolved with our cadence of 5 $t_g$.  However, given the lack of difference in $\Omega_{QU}$ when the observational data are smoothed by a 5 $t_g$ boxcar filter, we believe we are sufficiently capturing the loops that appear in the observations. 

\section{Scoring Consistent with Previous EHT Studies}
\label{sec:app_paperVIII_scoring}

\begin{figure*}
    \centering
    \includegraphics[width=\textwidth]{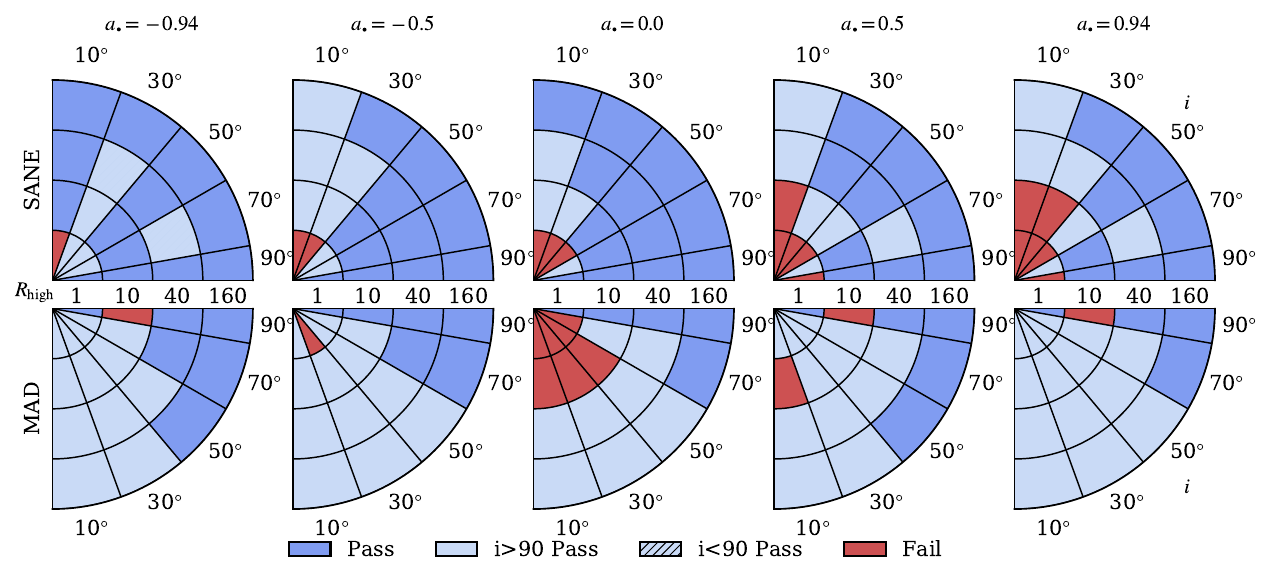} 
    \caption{As \autoref{fig:pizza}, but passing models if there is overlap if a model's 90 percent confidence region overlaps with the 90 percent confidence region of the data.  This more permissive scoring metric is consistent with previous EHT studies \citep{EHTC+2024c}.  More models pass than in \autoref{fig:pizza} due to the substantial scatter in $\Omega_{QU}$ and $f_{CW}$.}
    \label{fig:pizza_90}
\end{figure*}

In \autoref{fig:pizza}, we visualized the models whose 1$\sigma$ regions overlapped with the 1$\sigma$ regions of the data in $\Omega_{QU}$ and $f_{CW}$.  In previous EHT studies, a more permissive 90 percent confidence region overlap was used \citep{EHTC+2024c}.  In \autoref{fig:pizza_90}, we plot the results of this more permissive cut, finding a much larger fraction of passing models.  Far more models are able to pass with $i<90^\circ$ (either dark blue or hatched light blue), due simply to the significant time variability of both measured quantities.  This underscores the importance of continued monitoring of \sgra to reduce the observational error bar.

\bibliography{ms}

\begin{thebibliography}{}
\expandafter\ifx\csname natexlab\endcsname\relax\def\natexlab#1{#1}\fi
\providecommand{\url}[1]{\href{#1}{#1}}
\providecommand{\dodoi}[1]{doi:~\href{http://doi.org/#1}{\nolinkurl{#1}}}
\providecommand{\doeprint}[1]{\href{http://ascl.net/#1}{\nolinkurl{http://ascl.net/#1}}}
\providecommand{\doarXiv}[1]{\href{https://arxiv.org/abs/#1}{\nolinkurl{https://arxiv.org/abs/#1}}}

\bibitem[{N. {Aimar} {et~al.}(2023){Aimar}, {Dmytriiev}, {Vincent}, {El
  Mellah}, {Paumard}, {Perrin}, \& {Zech}}]{Aimar+2023}
{Aimar}, N., {Dmytriiev}, A., {Vincent}, F.~H., {et~al.} 2023,
  \bibinfo{title}{{Magnetic reconnection plasmoid model for Sagittarius A*
  flares},} \aap, 672, A62, \dodoi{10.1051/0004-6361/202244936}

\bibitem[{E. {Antonopoulou} {et~al.}(2025){Antonopoulou}, {Loules}, \&
  {Nathanail}}]{Antonopoulou+2025}
{Antonopoulou}, E., {Loules}, A., \& {Nathanail}, A. 2025,
  \bibinfo{title}{{Magnetically arrested disk flux eruption events to describe
  SgrA* flares},} arXiv e-prints, arXiv:2501.07521,
  \dodoi{10.48550/arXiv.2501.07521}

\bibitem[{B. {Balick} \& R.~L. {Brown}(1974){Balick} \&
  {Brown}}]{Balick&Brown1974}
{Balick}, B., \& {Brown}, R.~L. 1974, \bibinfo{title}{{Intense sub-arcsecond
  structure in the galactic center.},} \apj, 194, 265, \dodoi{10.1086/153242}

\bibitem[{E. {Berti} \& M. {Volonteri}(2008){Berti} \&
  {Volonteri}}]{Berti&Volonteri2008}
{Berti}, E., \& {Volonteri}, M. 2008, \bibinfo{title}{{Cosmological Black Hole
  Spin Evolution by Mergers and Accretion},} \apj, 684, 822,
  \dodoi{10.1086/590379}

\bibitem[{G.~S. {Bisnovatyi-Kogan} \& A.~A. {Ruzmaikin}(1974){Bisnovatyi-Kogan}
  \& {Ruzmaikin}}]{Bisnovatyi-Kogan&Ruzmaikin1974}
{Bisnovatyi-Kogan}, G.~S., \& {Ruzmaikin}, A.~A. 1974, \bibinfo{title}{{The
  Accretion of Matter by a Collapsing Star in the Presence of a Magnetic
  Field},} \apss, 28, 45, \dodoi{10.1007/BF00642237}

\bibitem[{G.~C. {Bower} {et~al.}(2015){Bower}, {Dexter}, {Markoff}, {Gurwell},
  {Rao}, \& {McHardy}}]{Bower+2015}
{Bower}, G.~C., {Dexter}, J., {Markoff}, S., {et~al.} 2015, \bibinfo{title}{{A
  Black Hole Mass-Variability Timescale Correlation at Submillimeter
  Wavelengths},} \apjl, 811, L6, \dodoi{10.1088/2041-8205/811/1/L6}

\bibitem[{G.~C. {Bower} {et~al.}(2005){Bower}, {Falcke}, {Wright}, \&
  {Backer}}]{Bower+2005}
{Bower}, G.~C., {Falcke}, H., {Wright}, M.~C., \& {Backer}, D.~C. 2005,
  \bibinfo{title}{{Variable Linear Polarization from Sagittarius A*: Evidence
  of a Hot Turbulent Accretion Flow},} \apjl, 618, L29, \dodoi{10.1086/427498}

\bibitem[{G.~C. {Bower} {et~al.}(2018){Bower}, {Broderick}, {Dexter},
  {Doeleman}, {Falcke}, {Fish}, {Johnson}, {Marrone}, {Moran}, {Moscibrodzka},
  {Peck}, {Plambeck}, \& {Rao}}]{Bower+2018}
{Bower}, G.~C., {Broderick}, A., {Dexter}, J., {et~al.} 2018,
  \bibinfo{title}{{ALMA Polarimetry of Sgr A*: Probing the Accretion Flow from
  the Event Horizon to the Bondi Radius},} \apj, 868, 101,
  \dodoi{10.3847/1538-4357/aae983}

\bibitem[{A.~E. {Broderick} \& A. {Loeb}(2006){Broderick} \&
  {Loeb}}]{Broderick&Loeb2006}
{Broderick}, A.~E., \& {Loeb}, A. 2006, \bibinfo{title}{{Imaging optically-thin
  hotspots near the black hole horizon of Sgr A* at radio and near-infrared
  wavelengths},} \mnras, 367, 905, \dodoi{10.1111/j.1365-2966.2006.10152.x}

\bibitem[{A.~E. {Broderick} {et~al.}(2022){Broderick}, {Gold}, {Georgiev},
  {Pesce}, {Tiede}, {Ni}, {Moriyama}, {Akiyama}, {Alberdi}, {Alef}, {Algaba},
  {Anantua}, {Asada}, {Azulay}, {Bach}, {Baczko}, {Ball}, {Balokovi{\'c}},
  {Barrett}, {Baub{\"o}ck}, {Benson}, {Bintley}, {Blackburn}, {Blundell},
  {Bouman}, {Bower}, {Boyce}, {Bremer}, {Brinkerink}, {Brissenden}, {Britzen},
  {Broguiere}, {Bronzwaer}, {Bustamante}, {Byun}, {Carlstrom}, {Ceccobello},
  {Chael}, {Chan}, {Chatterjee}, {Chatterjee}, {Chen}, {Chen}, {Cheng}, {Cho},
  {Christian}, {Conroy}, {Conway}, {Cordes}, {Crawford}, {Crew}, {Cruz-Osorio},
  {Cui}, {Davelaar}, {De Laurentis}, {Deane}, {Dempsey}, {Desvignes}, {Dexter},
  {Dhruv}, {Doeleman}, {Dougal}, {Dzib}, {Eatough}, {Emami}, {Falcke}, {Farah},
  {Fish}, {Fomalont}, {Ford}, {Fraga-Encinas}, {Freeman}, {Friberg}, {Fromm},
  {Fuentes}, {Galison}, {Gammie}, {Garc{\'\i}a}, {Gentaz}, {Goddi},
  {G{\'o}mez-Ruiz}, {G{\'o}mez}, {Gu}, {Gurwell}, {Hada}, {Haggard}, {Haworth},
  {Hecht}, {Hesper}, {Heumann}, {Ho}, {Ho}, {Honma}, {Huang}, {Huang},
  {Hughes}, {Ikeda}, {Impellizzeri}, {Inoue}, {Issaoun}, {James}, {Jannuzi},
  {Janssen}, {Jeter}, {Jiang}, {Jim{\'e}nez-Rosales}, {Johnson}, {Jorstad},
  {Joshi}, {Jung}, {Karami}, {Karuppusamy}, {Kawashima}, {Keating}, {Kettenis},
  {Kim}, {Kim}, {Kim}, {Kim}, {Kino}, {Koay}, {Kocherlakota}, {Kofuji}, {Koch},
  {Koyama}, {Kramer}, {Kramer}, {Krichbaum}, {Kuo}, {La Bella}, {Lauer}, {Lee},
  {Lee}, {Leung}, {Levis}, {Li}, {Lico}, {Lindahl}, {Lindqvist}, {Lisakov},
  {Liu}, {Liu}, {Liuzzo}, {Lo}, {Lobanov}, {Loinard}, {Lonsdale}, {Lu}, {Mao},
  {Marchili}, {Markoff}, {Marrone}, {Marscher}, {Mart{\'\i}-Vidal},
  {Matsushita}, {Matthews}, {Menten}, {Michalik}, {Mizuno}, {Mizuno}, {Moran},
  {Moscibrodzka}, {M{\"u}ller}, {Mus}, {Musoke}, {Myserlis}, {Nadolski},
  {Nagai}, {Nagar}, {Nakamura}, {Narayan}, {Narayanan}, {Natarajan},
  {Nathanail}, {Navarro Fuentes}, {Neilsen}, {Neri}, {Noutsos}, {Nowak}, {Oh},
  {Okino}, {Olivares}, {Ortiz-Le{\'o}n}, {Oyama}, {Palumbo}, {Paraschos},
  {Park}, {Parsons}, {Patel}, {Pen}, {Pi{\'e}tu}, {Plambeck}, {PopStefanija},
  {Porth}, {P{\"o}tzl}, {Prather}, {Preciado-L{\'o}pez}, {Pu}, {Ramakrishnan},
  \& {Rao}}]{Broderick+2022}
{Broderick}, A.~E., {Gold}, R., {Georgiev}, B., {et~al.} 2022,
  \bibinfo{title}{{Characterizing and Mitigating Intraday Variability:
  Reconstructing Source Structure in Accreting Black Holes with mm-VLBI},}
  \apjl, 930, L21, \dodoi{10.3847/2041-8213/ac6584}

\bibitem[{A. {Chael} {et~al.}(2023){Chael}, {Lupsasca}, {Wong}, \&
  {Quataert}}]{Chael+2023}
{Chael}, A., {Lupsasca}, A., {Wong}, G.~N., \& {Quataert}, E. 2023,
  \bibinfo{title}{{Black Hole Polarimetry I: A Signature of Electromagnetic
  Energy Extraction},} arXiv e-prints, arXiv:2307.06372,
  \dodoi{10.48550/arXiv.2307.06372}

\bibitem[{K. {Chatterjee} {et~al.}(2020){Chatterjee}, {Younsi}, {Liska},
  {Tchekhovskoy}, {Markoff}, {Yoon}, {van Eijnatten}, {Hesp}, {Ingram}, \& {van
  der Klis}}]{Chatterjee+2020}
{Chatterjee}, K., {Younsi}, Z., {Liska}, M., {et~al.} 2020,
  \bibinfo{title}{{Observational signatures of disc and jet misalignment in
  images of accreting black holes},} \mnras, 499, 362,
  \dodoi{10.1093/mnras/staa2718}

\bibitem[{K. {Chatterjee} {et~al.}(2021){Chatterjee}, {Markoff}, {Neilsen},
  {Younsi}, {Witzel}, {Tchekhovskoy}, {Yoon}, {Ingram}, {van der Klis},
  {Boyce}, {Do}, {Haggard}, \& {Nowak}}]{Chatterjee+2021}
{Chatterjee}, K., {Markoff}, S., {Neilsen}, J., {et~al.} 2021,
  \bibinfo{title}{{General relativistic MHD simulations of non-thermal flaring
  in Sagittarius A*},} \mnras, 507, 5281, \dodoi{10.1093/mnras/stab2466}

\bibitem[{B.-Y. {Chen} {et~al.}(2023){Chen}, {Bower}, {Dexter}, {Markoff},
  {Ridenour}, {Gurwell}, {Rao}, \& {Wallstr{\"o}m}}]{Chen+2023}
{Chen}, B.-Y., {Bower}, G.~C., {Dexter}, J., {et~al.} 2023,
  \bibinfo{title}{{Testing the Linear Relationship between Black Hole Mass and
  Variability Timescale in Low-luminosity AGNs at Submillimeter Wavelengths},}
  \apj, 951, 93, \dodoi{10.3847/1538-4357/acd250}

\bibitem[{N.~S. {Conroy} {et~al.}(2023){Conroy}, {Baub{\"o}ck}, {Dhruv}, {Lee},
  {Broderick}, {Chan}, {Georgiev}, {Joshi}, {Prather}, \&
  {Gammie}}]{Conroy+2023}
{Conroy}, N.~S., {Baub{\"o}ck}, M., {Dhruv}, V., {et~al.} 2023,
  \bibinfo{title}{{Rotation in Event Horizon Telescope Movies},} \apj, 951, 46,
  \dodoi{10.3847/1538-4357/acd2c8}

\bibitem[{J.-P. {De Villiers} {et~al.}(2003){De Villiers}, {Hawley}, \&
  {Krolik}}]{DeVilliers+2003}
{De Villiers}, J.-P., {Hawley}, J.~F., \& {Krolik}, J.~H. 2003,
  \bibinfo{title}{{Magnetically Driven Accretion Flows in the Kerr Metric. I.
  Models and Overall Structure},} \apj, 599, 1238, \dodoi{10.1086/379509}

\bibitem[{J. {Dexter} {et~al.}(2014){Dexter}, {Kelly}, {Bower}, {Marrone},
  {Stone}, \& {Plambeck}}]{Dexter+2014}
{Dexter}, J., {Kelly}, B., {Bower}, G.~C., {et~al.} 2014, \bibinfo{title}{{An 8
  h characteristic time-scale in submillimetre light curves of Sagittarius
  A*},} \mnras, 442, 2797, \dodoi{10.1093/mnras/stu1039}

\bibitem[{J. {Dexter} {et~al.}(2020){Dexter}, {Tchekhovskoy},
  {Jim{\'e}nez-Rosales}, {Ressler}, {Baub{\"o}ck}, {Dallilar}, {de Zeeuw},
  {Eisenhauer}, {von Fellenberg}, {Gao}, {Genzel}, {Gillessen}, {Habibi},
  {Ott}, {Stadler}, {Straub}, \& {Widmann}}]{Dexter+2020}
{Dexter}, J., {Tchekhovskoy}, A., {Jim{\'e}nez-Rosales}, A., {et~al.} 2020,
  \bibinfo{title}{{Sgr A* near-infrared flares from reconnection events in a
  magnetically arrested disc},} \mnras, 497, 4999,
  \dodoi{10.1093/mnras/staa2288}

\bibitem[{V. {Dhruv} {et~al.}(2025){Dhruv}, {Prather}, {Wong}, \&
  {Gammie}}]{Dhruv+2025}
{Dhruv}, V., {Prather}, B., {Wong}, G.~N., \& {Gammie}, C.~F. 2025,
  \bibinfo{title}{{A Survey of General Relativistic Magnetohydrodynamic Models
  for Black Hole Accretion Systems},} \apjs, 277, 16,
  \dodoi{10.3847/1538-4365/adaea6}

\bibitem[{T. {Do} {et~al.}(2019){Do}, {Hees}, {Ghez}, {Martinez}, {Chu}, {Jia},
  {Sakai}, {Lu}, {Gautam}, {O'Neil}, {Becklin}, {Morris}, {Matthews},
  {Nishiyama}, {Campbell}, {Chappell}, {Chen}, {Ciurlo}, {Dehghanfar},
  {Gallego-Cano}, {Kerzendorf}, {Lyke}, {Naoz}, {Saida}, {Sch{\"o}del},
  {Takahashi}, {Takamori}, {Witzel}, \& {Wizinowich}}]{Do+2019}
{Do}, T., {Hees}, A., {Ghez}, A., {et~al.} 2019, \bibinfo{title}{{Relativistic
  redshift of the star S0-2 orbiting the Galactic Center supermassive black
  hole},} Science, 365, 664, \dodoi{10.1126/science.aav8137}

\bibitem[{I. {El Mellah} {et~al.}(2023){El Mellah}, {Cerutti}, \&
  {Crinquand}}]{ElMellah+2023}
{El Mellah}, I., {Cerutti}, B., \& {Crinquand}, B. 2023,
  \bibinfo{title}{{Reconnection-driven flares in 3D black hole magnetospheres.
  A scenario for hot spots around Sagittarius A*},} \aap, 677, A67,
  \dodoi{10.1051/0004-6361/202346781}

\bibitem[{R. {Emami} {et~al.}(2023{\natexlab{a}}){Emami}, {Doeleman},
  {Wielgus}, {Chang}, {Chatterjee}, {Smith}, {Liska}, {Steiner}, {Ricarte},
  {Narayan}, {Tremblay}, {Finkbeiner}, {Hernquist}, {Chan}, {Blackburn},
  {Prather}, {Tiede}, {Broderick}, {Vogelsberger}, {Alcock}, \&
  {Roelofs}}]{Emami+2023d}
{Emami}, R., {Doeleman}, S.~S., {Wielgus}, M., {et~al.} 2023{\natexlab{a}},
  \bibinfo{title}{{The EB Correlation in Resolved Polarized Images: Connections
  to the Astrophysics of Black Holes},} \apj, 955, 6,
  \dodoi{10.3847/1538-4357/acdc96}

\bibitem[{R. {Emami} {et~al.}(2023{\natexlab{b}}){Emami}, {Ricarte}, {Wong},
  {Palumbo}, {Chang}, {Doeleman}, {Broderick}, {Narayan}, {Wielgus},
  {Blackburn}, {Prather}, {Chael}, {Anantua}, {Chatterjee}, {Marti-Vidal},
  {G{\'o}mez}, {Akiyama}, {Liska}, {Hernquist}, {Tremblay}, {Vogelsberger},
  {Alcock}, {Smith}, {Steiner}, {Tiede}, \& {Roelofs}}]{Emami+2023}
{Emami}, R., {Ricarte}, A., {Wong}, G.~N., {et~al.} 2023{\natexlab{b}},
  \bibinfo{title}{{Unraveling Twisty Linear Polarization Morphologies in Black
  Hole Images},} \apj, 950, 38, \dodoi{10.3847/1538-4357/acc8cd}

\bibitem[{ {Event Horizon Telescope Collaboration} {et~al.}(2021){Event Horizon
  Telescope Collaboration}, {Akiyama}, {Algaba}, {Alberdi}, {Alef}, {Anantua},
  {Asada}, {Azulay}, {Baczko}, {Ball}, {Balokovi{\'c}}, {Barrett}, {Benson},
  {Bintley}, {Blackburn}, {Blundell}, {Boland}, {Bouman}, {Bower}, {Boyce},
  {Bremer}, {Brinkerink}, {Brissenden}, {Britzen}, {Broderick}, {Broguiere},
  {Bronzwaer}, {Byun}, {Carlstrom}, {Chael}, {Chan}, {Chatterjee},
  {Chatterjee}, {Chen}, {Chen}, {Chesler}, {Cho}, {Christian}, {Conway},
  {Cordes}, {Crawford}, {Crew}, {Cruz-Osorio}, {Cui}, {Davelaar}, {De
  Laurentis}, {Deane}, {Dempsey}, {Desvignes}, {Dexter}, {Doeleman}, {Eatough},
  {Falcke}, {Farah}, {Fish}, {Fomalont}, {Ford}, {Fraga-Encinas}, {Friberg},
  {Fromm}, {Fuentes}, {Galison}, {Gammie}, {Garc{\'\i}a}, {Gelles}, {Gentaz},
  {Georgiev}, {Goddi}, {Gold}, {G{\'o}mez}, {G{\'o}mez-Ruiz}, {Gu}, {Gurwell},
  {Hada}, {Haggard}, {Hecht}, {Hesper}, {Himwich}, {Ho}, {Ho}, {Honma},
  {Huang}, {Huang}, {Hughes}, {Ikeda}, {Inoue}, {Issaoun}, {James}, {Jannuzi},
  {Janssen}, {Jeter}, {Jiang}, {Jimenez-Rosales}, {Johnson}, {Jorstad}, {Jung},
  {Karami}, {Karuppusamy}, {Kawashima}, {Keating}, {Kettenis}, {Kim}, {Kim},
  {Kim}, {Kim}, {Kino}, {Koay}, {Kofuji}, {Koch}, {Koyama}, {Kramer}, {Kramer},
  {Krichbaum}, {Kuo}, {Lauer}, {Lee}, {Levis}, {Li}, {Li}, {Lindqvist}, {Lico},
  {Lindahl}, {Liu}, {Liu}, {Liuzzo}, {Lo}, {Lobanov}, {Loinard}, {Lonsdale},
  {Lu}, {MacDonald}, {Mao}, {Marchili}, {Markoff}, {Marrone}, {Marscher},
  {Mart{\'\i}-Vidal}, {Matsushita}, {Matthews}, {Medeiros}, {Menten}, {Mizuno},
  {Mizuno}, {Moran}, {Moriyama}, {Moscibrodzka}, {M{\"u}ller}, {Musoke}, {Mus
  Mej{\'\i}as}, {Michalik}, {Nadolski}, {Nagai}, {Nagar}, {Nakamura},
  {Narayan}, {Narayanan}, {Natarajan}, {Nathanail}, {Neilsen}, {Neri}, {Ni},
  {Noutsos}, {Nowak}, {Okino}, {Olivares}, {Ortiz-Le{\'o}n}, {Oyama},
  {{\"O}zel}, {Palumbo}, {Park}, {Patel}, {Pen}, {Pesce}, {Pi{\'e}tu},
  {Plambeck}, {PopStefanija}, {Porth}, {P{\"o}tzl}, {Prather},
  {Preciado-L{\'o}pez}, {Psaltis}, {Pu}, {Ramakrishnan}, {Rao}, {Rawlings},
  {Raymond}, {Rezzolla}, {Ricarte}, {Ripperda}, {Roelofs}, {Rogers}, {Ros},
  {Rose}, {Roshanineshat}, {Rottmann}, {Roy}, {Ruszczyk}, {Rygl},
  {S{\'a}nchez}, {S{\'a}nchez-Arguelles}, {Sasada}, {Savolainen}, {Schloerb},
  {Schuster}, {Shao}, {Shen}, {Small}, {Sohn}, {SooHoo}, {Sun}, {Tazaki},
  {Tetarenko}, {Tiede}, {Tilanus}, {Titus}, {Toma}, {Torne}, {Trent},
  {Traianou}, {Trippe}, {van Bemmel}, {van Langevelde}, {van Rossum}, {Wagner},
  {Ward-Thompson}, {Wardle}, {Weintroub}, {Wex}, {Wharton}, {Wielgus}, {Wong},
  {Wu}, {Yoon}, {Young}, {Young}, {Younsi}, {Yuan}, {Yuan}, {Zensus}, {Zhao},
  \& {Zhao}}]{EHTC+2021b}
{Event Horizon Telescope Collaboration}, {Akiyama}, K., {Algaba}, J.~C.,
  {et~al.} 2021, \bibinfo{title}{{First M87 Event Horizon Telescope Results.
  VIII. Magnetic Field Structure near The Event Horizon},} \apjl, 910, L13,
  \dodoi{10.3847/2041-8213/abe4de}

\bibitem[{ {Event Horizon Telescope Collaboration}
  {et~al.}(2022{\natexlab{a}}){Event Horizon Telescope Collaboration},
  {Akiyama}, {Alberdi}, {Alef}, {Algaba}, {Anantua}, {Asada}, {Azulay}, {Bach},
  {Baczko}, {Ball}, {Balokovi{\'c}}, {Barrett}, {Baub{\"o}ck}, {Benson},
  {Bintley}, {Blackburn}, {Blundell}, {Bouman}, {Bower}, {Boyce}, {Bremer},
  {Brinkerink}, {Brissenden}, {Britzen}, {Broderick}, {Broguiere}, {Bronzwaer},
  {Bustamante}, {Byun}, {Carlstrom}, {Ceccobello}, {Chael}, {Chan},
  {Chatterjee}, {Chatterjee}, {Chen}, {Chen}, {Cheng}, {Cho}, {Christian},
  {Conroy}, {Conway}, {Cordes}, {Crawford}, {Crew}, {Cruz-Osorio}, {Cui},
  {Davelaar}, {De Laurentis}, {Deane}, {Dempsey}, {Desvignes}, {Dexter},
  {Dhruv}, {Doeleman}, {Dougal}, {Dzib}, {Eatough}, {Emami}, {Falcke}, {Farah},
  {Fish}, {Fomalont}, {Ford}, {Fraga-Encinas}, {Freeman}, {Friberg}, {Fromm},
  {Fuentes}, {Galison}, {Gammie}, {Garc{\'\i}a}, {Gentaz}, {Georgiev}, {Goddi},
  {Gold}, {G{\'o}mez-Ruiz}, {G{\'o}mez}, {Gu}, {Gurwell}, {Hada}, {Haggard},
  {Haworth}, {Hecht}, {Hesper}, {Heumann}, {Ho}, {Ho}, {Honma}, {Huang},
  {Huang}, {Hughes}, {Ikeda}, {Impellizzeri}, {Inoue}, {Issaoun}, {James},
  {Jannuzi}, {Janssen}, {Jeter}, {Jiang}, {Jim{\'e}nez-Rosales}, {Johnson},
  {Jorstad}, {Joshi}, {Jung}, {Karami}, {Karuppusamy}, {Kawashima}, {Keating},
  {Kettenis}, {Kim}, {Kim}, {Kim}, {Kim}, {Kino}, {Koay}, {Kocherlakota},
  {Kofuji}, {Koch}, {Koyama}, {Kramer}, {Kramer}, {Krichbaum}, {Kuo}, {La
  Bella}, {Lauer}, {Lee}, {Lee}, {Leung}, {Levis}, {Li}, {Lico}, {Lindahl},
  {Lindqvist}, {Lisakov}, {Liu}, {Liu}, {Liuzzo}, {Lo}, {Lobanov}, {Loinard},
  {Lonsdale}, {Lu}, {Mao}, {Marchili}, {Markoff}, {Marrone}, {Marscher},
  {Mart{\'\i}-Vidal}, {Matsushita}, {Matthews}, {Medeiros}, {Menten},
  {Michalik}, {Mizuno}, {Mizuno}, {Moran}, {Moriyama}, {Moscibrodzka},
  {M{\"u}ller}, {Mus}, {Musoke}, {Myserlis}, {Nadolski}, {Nagai}, {Nagar},
  {Nakamura}, {Narayan}, {Narayanan}, {Natarajan}, {Nathanail}, {Navarro
  Fuentes}, {Neilsen}, {Neri}, {Ni}, {Noutsos}, {Nowak}, {Oh}, {Okino},
  {Olivares}, {Ortiz-Le{\'o}n}, {Oyama}, {{\"O}zel}, {Palumbo}, {Paraschos},
  {Park}, {Parsons}, {Patel}, {Pen}, {Pesce}, {Pi{\'e}tu}, {Plambeck},
  {PopStefanija}, {Porth}, {P{\"o}tzl}, {Prather}, {Preciado-L{\'o}pez},
  {Psaltis}, {Pu}, {Ramakrishnan}, {Rao}, {Rawlings}, {Raymond}, {Rezzolla},
  {Ricarte}, {Ripperda}, {Roelofs}, {Rogers}, {Ros}, {Romero-Ca{\~n}izales},
  {Roshanineshat}, {Rottmann}, {Roy}, {Ruiz}, {Ruszczyk}, {Rygl},
  {S{\'a}nchez}, {S{\'a}nchez-Arg{\"u}elles}, {S{\'a}nchez-Portal}, {Sasada},
  {Satapathy}, {Savolainen}, {Schloerb}, {Schonfeld}, {Schuster}, {Shao},
  {Shen}, {Small}, {Sohn}, {SooHoo}, {Souccar}, {Sun}, {Tazaki}, {Tetarenko},
  {Tiede}, {Tilanus}, {Titus}, {Torne}, {Traianou}, {Trent}, {Trippe}, {Turk},
  {van Bemmel}, {van Langevelde}, {van Rossum}, {Vos}, {Wagner},
  {Ward-Thompson}, {Wardle}, {Weintroub}, {Wex}, {Wharton}, {Wielgus}, {Wiik},
  {Witzel}, {Wondrak}, {Wong}, {Wu}, {Yamaguchi}, {Yoon}, {Young}, {Young},
  {Younsi}, {Yuan}, {Yuan}, {Zensus}, {Zhang}, {Zhao}, {Zhao}, {Agurto},
  {Allardi}, {Amestica}, {Araneda}, {Arriagada}, {Berghuis}, {Bertarini},
  {Berthold}, {Blanchard}, {Brown}, {C{\'a}rdenas}, {Cantzler}, {Caro},
  {Castillo-Dom{\'\i}nguez}, {Chan}, {Chang}, {Chang}, {Chang}, {Chang},
  {Chen}, {Chilson}, {Chuter}, {Ciechanowicz}, {Colin-Beltran}, {Coulson},
  {Crowley}, {Degenaar}, {Dornbusch}, {Dur{\'a}n}, {Everett}, {Faber},
  {Forster}, {Fuchs}, {Gale}, {Geertsema}, {Gonz{\'a}lez}, {Graham}, {Gueth},
  {Halverson}, {Han}, {Han}, {Hasegawa}, {Hern{\'a}ndez-Rebollar}, {Herrera},
  {Herrero-Illana}, {Heyminck}, {Hirota}, {Hoge}, {Hostler Schimpf}, {Howie},
  {Huang}, {Jiang}, {Jinchi}, {John}, {Kimura}, {Klein}, {Kubo}, {Kuroda},
  {Kwon}, {Lacasse}, {Laing}, {Leitch}, {Li}, {Liu}, {Liu}, {Lin}, {Lu},
  {Mac-Auliffe}, {Martin-Cocher}, {Matulonis}, {Maute}, {Messias},
  {Meyer-Zhao}, {Monta{\~n}a}, {Montenegro-Montes}, {Montgomerie}, {Moreno
  Nolasco}, {Muders}, {Nishioka}, {Norton}, {Nystrom}, {Ogawa}, {Olivares},
  {Oshiro}, {P{\'e}rez-Beaupuits}, {Parra}, {Phillips}, {Poirier}, {Pradel},
  {Qiu}, {Raffin}, {Rahlin}, {Ram{\'\i}rez}, {Ressler}, {Reynolds},
  {Rodr{\'\i}guez-Montoya}, {Saez-Madain}, {Santana}, {Shaw}, {Shirkey},
  {Silva}, {Snow}, {Sousa}, {Sridharan}, {Stahm}, {Stark}, {Test},
  {Torstensson}, {Venegas}, {Walther}, {Wei}, {White}, {Wieching}, {Wijnands},
  {Wouterloot}, {Yu}, {Yu}, {Zeballos}, \& {EHT Collaboration}}]{EHTC+2022a}
{Event Horizon Telescope Collaboration}, {Akiyama}, K., {Alberdi}, A., {et~al.}
  2022{\natexlab{a}}, \bibinfo{title}{{First Sagittarius A* Event Horizon
  Telescope Results. I. The Shadow of the Supermassive Black Hole in the Center
  of the Milky Way},} \apjl, 930, L12, \dodoi{10.3847/2041-8213/ac6674}

\bibitem[{ {Event Horizon Telescope Collaboration}
  {et~al.}(2022{\natexlab{b}}){Event Horizon Telescope Collaboration},
  {Akiyama}, {Alberdi}, {Alef}, {Algaba}, {Anantua}, {Asada}, {Azulay}, {Bach},
  {Baczko}, {Ball}, {Balokovi{\'c}}, {Barrett}, {Baub{\"o}ck}, {Benson},
  {Bintley}, {Blackburn}, {Blundell}, {Bouman}, {Bower}, {Boyce}, {Bremer},
  {Brinkerink}, {Brissenden}, {Britzen}, {Broderick}, {Broguiere}, {Bronzwaer},
  {Bustamante}, {Byun}, {Carlstrom}, {Ceccobello}, {Chael}, {Chan},
  {Chatterjee}, {Chatterjee}, {Chen}, {Chen}, {Cheng}, {Cho}, {Christian},
  {Conroy}, {Conway}, {Cordes}, {Crawford}, {Crew}, {Cruz-Osorio}, {Cui},
  {Davelaar}, {De Laurentis}, {Deane}, {Dempsey}, {Desvignes}, {Dexter},
  {Dhruv}, {Doeleman}, {Dougal}, {Dzib}, {Eatough}, {Emami}, {Falcke}, {Farah},
  {Fish}, {Fomalont}, {Ford}, {Fraga-Encinas}, {Freeman}, {Friberg}, {Fromm},
  {Fuentes}, {Galison}, {Gammie}, {Garc{\'\i}a}, {Gentaz}, {Georgiev}, {Goddi},
  {Gold}, {G{\'o}mez-Ruiz}, {G{\'o}mez}, {Gu}, {Gurwell}, {Hada}, {Haggard},
  {Haworth}, {Hecht}, {Hesper}, {Heumann}, {Ho}, {Ho}, {Honma}, {Huang},
  {Huang}, {Hughes}, {Ikeda}, {Impellizzeri}, {Inoue}, {Issaoun}, {James},
  {Jannuzi}, {Janssen}, {Jeter}, {Jiang}, {Jim{\'e}nez-Rosales}, {Johnson},
  {Jorstad}, {Joshi}, {Jung}, {Karami}, {Karuppusamy}, {Kawashima}, {Keating},
  {Kettenis}, {Kim}, {Kim}, {Kim}, {Kim}, {Kino}, {Koay}, {Kocherlakota},
  {Kofuji}, {Koch}, {Koyama}, {Kramer}, {Kramer}, {Krichbaum}, {Kuo}, {La
  Bella}, {Lauer}, {Lee}, {Lee}, {Leung}, {Levis}, {Li}, {Lico}, {Lindahl},
  {Lindqvist}, {Lisakov}, {Liu}, {Liu}, {Liuzzo}, {Lo}, {Lobanov}, {Loinard},
  {Lonsdale}, {Lu}, {Mao}, {Marchili}, {Markoff}, {Marrone}, {Marscher},
  {Mart{\'\i}-Vidal}, {Matsushita}, {Matthews}, {Medeiros}, {Menten},
  {Michalik}, {Mizuno}, {Mizuno}, {Moran}, {Moriyama}, {Moscibrodzka},
  {M{\"u}ller}, {Mus}, {Musoke}, {Myserlis}, {Nadolski}, {Nagai}, {Nagar},
  {Nakamura}, {Narayan}, {Narayanan}, {Natarajan}, {Nathanail}, {Navarro
  Fuentes}, {Neilsen}, {Neri}, {Ni}, {Noutsos}, {Nowak}, {Oh}, {Okino},
  {Olivares}, {Ortiz-Le{\'o}n}, {Oyama}, {Palumbo}, {Paraschos}, {Park},
  {Parsons}, {Patel}, {Pen}, {Pesce}, {Pi{\'e}tu}, {Plambeck}, {PopStefanija},
  {Porth}, {P{\"o}tzl}, {Prather}, {Preciado-L{\'o}pez}, {Pu}, {Ramakrishnan},
  {Rao}, {Rawlings}, {Raymond}, {Rezzolla}, {Ricarte}, {Ripperda}, {Roelofs},
  {Rogers}, {Ros}, {Romero-Ca{\~n}izales}, {Roshanineshat}, {Rottmann}, {Roy},
  {Ruiz}, {Ruszczyk}, {Rygl}, {S{\'a}nchez}, {S{\'a}nchez-Arg{\"u}elles},
  {S{\'a}nchez-Portal}, {Sasada}, {Satapathy}, {Savolainen}, {Schloerb},
  {Schonfeld}, {Schuster}, {Shao}, {Shen}, {Small}, {Sohn}, {SooHoo},
  {Souccar}, {Sun}, {Tazaki}, {Tetarenko}, {Tiede}, {Tilanus}, {Titus},
  {Torne}, {Traianou}, {Trent}, {Trippe}, {Turk}, {van Bemmel}, {van
  Langevelde}, {van Rossum}, {Vos}, {Wagner}, {Ward-Thompson}, {Wardle},
  {Weintroub}, {Wex}, {Wharton}, {Wielgus}, {Wiik}, {Witzel}, {Wondrak},
  {Wong}, {Wu}, {Yamaguchi}, {Yoon}, {Young}, {Young}, {Younsi}, {Yuan},
  {Yuan}, {Zensus}, {Zhang}, {Zhao}, {Zhao}, {Chang}, \& {EHT
  Collaboration}}]{EHTC+2022d}
{Event Horizon Telescope Collaboration}, {Akiyama}, K., {Alberdi}, A., {et~al.}
  2022{\natexlab{b}}, \bibinfo{title}{{First Sagittarius A* Event Horizon
  Telescope Results. IV. Variability, Morphology, and Black Hole Mass},} \apjl,
  930, L15, \dodoi{10.3847/2041-8213/ac6736}

\bibitem[{ {Event Horizon Telescope Collaboration}
  {et~al.}(2022{\natexlab{c}}){Event Horizon Telescope Collaboration},
  {Akiyama}, {Alberdi}, {Alef}, {Algaba}, {Anantua}, {Asada}, {Azulay}, {Bach},
  {Baczko}, {Ball}, {Balokovi{\'c}}, {Barrett}, {Baub{\"o}ck}, {Benson},
  {Bintley}, {Blackburn}, {Blundell}, {Bouman}, {Bower}, {Boyce}, {Bremer},
  {Brinkerink}, {Brissenden}, {Britzen}, {Broderick}, {Broguiere}, {Bronzwaer},
  {Bustamante}, {Byun}, {Carlstrom}, {Ceccobello}, {Chael}, {Chan},
  {Chatterjee}, {Chatterjee}, {Chen}, {Chen}, {Cheng}, {Cho}, {Christian},
  {Conroy}, {Conway}, {Cordes}, {Crawford}, {Crew}, {Cruz-Osorio}, {Cui},
  {Davelaar}, {De Laurentis}, {Deane}, {Dempsey}, {Desvignes}, {Dexter},
  {Dhruv}, {Doeleman}, {Dougal}, {Dzib}, {Eatough}, {Emami}, {Falcke}, {Farah},
  {Fish}, {Fomalont}, {Ford}, {Fraga-Encinas}, {Freeman}, {Friberg}, {Fromm},
  {Fuentes}, {Galison}, {Gammie}, {Garc{\'\i}a}, {Gentaz}, {Georgiev}, {Goddi},
  {Gold}, {G{\'o}mez-Ruiz}, {G{\'o}mez}, {Gu}, {Gurwell}, {Hada}, {Haggard},
  {Haworth}, {Hecht}, {Hesper}, {Heumann}, {Ho}, {Ho}, {Honma}, {Huang},
  {Huang}, {Hughes}, {Ikeda}, {Violette Impellizzeri}, {Inoue}, {Issaoun},
  {James}, {Jannuzi}, {Janssen}, {Jeter}, {Jiang}, {Jim{\'e}nez-Rosales},
  {Johnson}, {Jorstad}, {Joshi}, {Jung}, {Karami}, {Karuppusamy}, {Kawashima},
  {Keating}, {Kettenis}, {Kim}, {Kim}, {Kim}, {Kim}, {Kino}, {Koay},
  {Kocherlakota}, {Kofuji}, {Koch}, {Koyama}, {Kramer}, {Kramer}, {Krichbaum},
  {Kuo}, {La Bella}, {Lauer}, {Lee}, {Lee}, {Leung}, {Levis}, {Li}, {Lico},
  {Lindahl}, {Lindqvist}, {Lisakov}, {Liu}, {Liu}, {Liuzzo}, {Lo}, {Lobanov},
  {Loinard}, {Lonsdale}, {Lu}, {Mao}, {Marchili}, {Markoff}, {Marrone},
  {Marscher}, {Mart{\'\i}-Vidal}, {Matsushita}, {Matthews}, {Medeiros},
  {Menten}, {Michalik}, {Mizuno}, {Mizuno}, {Moran}, {Moriyama},
  {Moscibrodzka}, {M{\"u}ller}, {Mus}, {Musoke}, {Myserlis}, {Nadolski},
  {Nagai}, {Nagar}, {Nakamura}, {Narayan}, {Narayanan}, {Natarajan},
  {Nathanail}, {Navarro Fuentes}, {Neilsen}, {Neri}, {Ni}, {Noutsos}, {Nowak},
  {Oh}, {Okino}, {Olivares}, {Ortiz-Le{\'o}n}, {Oyama}, {{\"O}zel}, {Palumbo},
  {Filippos Paraschos}, {Park}, {Parsons}, {Patel}, {Pen}, {Pesce},
  {Pi{\'e}tu}, {Plambeck}, {PopStefanija}, {Porth}, {P{\"o}tzl}, {Prather},
  {Preciado-L{\'o}pez}, {Psaltis}, {Pu}, {Ramakrishnan}, {Rao}, {Rawlings},
  {Raymond}, {Rezzolla}, {Ricarte}, {Ripperda}, {Roelofs}, {Rogers}, {Ros},
  {Romero-Ca{\~n}izales}, {Roshanineshat}, {Rottmann}, {Roy}, {Ruiz},
  {Ruszczyk}, {Rygl}, {S{\'a}nchez}, {S{\'a}nchez-Arg{\"u}elles},
  {S{\'a}nchez-Portal}, {Sasada}, {Satapathy}, {Savolainen}, {Schloerb},
  {Schonfeld}, {Schuster}, {Shao}, {Shen}, {Small}, {Sohn}, {SooHoo},
  {Souccar}, {Sun}, {Tazaki}, {Tetarenko}, {Tiede}, {Tilanus}, {Titus},
  {Torne}, {Traianou}, {Trent}, {Trippe}, {Turk}, {van Bemmel}, {van
  Langevelde}, {van Rossum}, {Vos}, {Wagner}, {Ward-Thompson}, {Wardle},
  {Weintroub}, {Wex}, {Wharton}, {Wielgus}, {Wiik}, {Witzel}, {Wondrak},
  {Wong}, {Wu}, {Yamaguchi}, {Yoon}, {Young}, {Young}, {Younsi}, {Yuan},
  {Yuan}, {Zensus}, {Zhang}, {Zhao}, {Zhao}, {Chan}, {Qiu}, {Ressler}, \&
  {White}}]{EHTC+2022e}
{Event Horizon Telescope Collaboration}, {Akiyama}, K., {Alberdi}, A., {et~al.}
  2022{\natexlab{c}}, \bibinfo{title}{{First Sagittarius A* Event Horizon
  Telescope Results. V. Testing Astrophysical Models of the Galactic Center
  Black Hole},} \apjl, 930, L16, \dodoi{10.3847/2041-8213/ac6672}

\bibitem[{ {Event Horizon Telescope Collaboration} {et~al.}(2023){Event Horizon
  Telescope Collaboration}, {Akiyama}, {Alberdi}, {Alef}, {Algaba}, {Anantua},
  {Asada}, {Azulay}, {Bach}, {Baczko}, {Ball}, {Balokovi{\'c}}, {Barrett},
  {Baub{\"o}ck}, {Benson}, {Bintley}, {Blackburn}, {Blundell}, {Bouman},
  {Bower}, {Boyce}, {Bremer}, {Brinkerink}, {Brissenden}, {Britzen},
  {Broderick}, {Broguiere}, {Bronzwaer}, {Bustamante}, {Byun}, {Carlstrom},
  {Ceccobello}, {Chael}, {Chan}, {Chang}, {Chatterjee}, {Chatterjee}, {Chen},
  {Chen}, {Cheng}, {Cho}, {Christian}, {Conroy}, {Conway}, {Cordes},
  {Crawford}, {Crew}, {Cruz-Osorio}, {Cui}, {Dahale}, {Davelaar}, {De
  Laurentis}, {Deane}, {Dempsey}, {Desvignes}, {Dexter}, {Dhruv}, {Doeleman},
  {Dougal}, {Dzib}, {Eatough}, {Emami}, {Falcke}, {Farah}, {Fish}, {Fomalont},
  {Ford}, {Foschi}, {Fraga-Encinas}, {Freeman}, {Friberg}, {Fromm}, {Fuentes},
  {Galison}, {Gammie}, {Garc{\'\i}a}, {Gentaz}, {Georgiev}, {Goddi}, {Gold},
  {G{\'o}mez-Ruiz}, {G{\'o}mez}, {Gu}, {Gurwell}, {Hada}, {Haggard}, {Haworth},
  {Hecht}, {Hesper}, {Heumann}, {Ho}, {Ho}, {Honma}, {Huang}, {Huang},
  {Hughes}, {Ikeda}, {Impellizzeri}, {Inoue}, {Issaoun}, {James}, {Jannuzi},
  {Janssen}, {Jeter}, {Jiang}, {Jim{\'e}nez-Rosales}, {Johnson}, {Jorstad},
  {Joshi}, {Jung}, {Karami}, {Karuppusamy}, {Kawashima}, {Keating}, {Kettenis},
  {Kim}, {Kim}, {Kim}, {Kim}, {Kino}, {Koay}, {Kocherlakota}, {Kofuji}, {Koch},
  {Koyama}, {Kramer}, {Kramer}, {Kramer}, {Krichbaum}, {Kuo}, {La Bella},
  {Lauer}, {Lee}, {Lee}, {Leung}, {Levis}, {Li}, {Lico}, {Lindahl},
  {Lindqvist}, {Lisakov}, {Liu}, {Liu}, {Liuzzo}, {Lo}, {Lobanov}, {Loinard},
  {Lonsdale}, {Lowitz}, {Lu}, {MacDonald}, {Mao}, {Marchili}, {Markoff},
  {Marrone}, {Marscher}, {Mart{\'\i}-Vidal}, {Matsushita}, {Matthews},
  {Medeiros}, {Menten}, {Michalik}, {Mizuno}, {Mizuno}, {Moran}, {Moriyama},
  {Moscibrodzka}, {Mulaudzi}, {M{\"u}ller}, {M{\"u}ller}, {Mus}, {Musoke},
  {Myserlis}, {Nadolski}, {Nagai}, {Nagar}, {Nakamura}, {Narayan}, {Narayanan},
  {Natarajan}, {Nathanail}, {Fuentes}, {Neilsen}, {Neri}, {Ni}, {Noutsos},
  {Nowak}, {Oh}, {Okino}, {Olivares}, {Ortiz-Le{\'o}n}, {Oyama}, {{\"O}zel},
  {Palumbo}, {Paraschos}, {Park}, {Parsons}, {Patel}, {Pen}, {Pesce},
  {Pi{\'e}tu}, {Plambeck}, {PopStefanija}, {Porth}, {P{\"o}tzl}, {Prather},
  {Preciado-L{\'o}pez}, {Psaltis}, {Pu}, {Ramakrishnan}, {Rao}, {Rawlings},
  {Raymond}, {Rezzolla}, {Ricarte}, {Ripperda}, {Roelofs}, {Rogers},
  {Romero-Ca{\~n}izales}, {Ros}, {Roshanineshat}, {Rottmann}, {Roy}, {Ruiz},
  {Ruszczyk}, {Rygl}, {S{\'a}nchez}, {S{\'a}nchez-Arg{\"u}elles},
  {S{\'a}nchez-Portal}, {Sasada}, {Satapathy}, {Savolainen}, {Schloerb},
  {Schonfeld}, {Schuster}, {Shao}, {Shen}, {Small}, {Sohn}, {SooHoo},
  {Sosapanta Salas}, {Souccar}, {Sun}, {Tazaki}, {Tetarenko}, {Tiede},
  {Tilanus}, {Titus}, {Torne}, {Toscano}, {Traianou}, {Trent}, {Trippe},
  {Turk}, {van Bemmel}, {van Langevelde}, {van Rossum}, {Vos}, {Wagner},
  {Ward-Thompson}, {Wardle}, {Washington}, {Weintroub}, {Wharton}, {Wielgus},
  {Wiik}, {Witzel}, {Wondrak}, {Wong}, {Wu}, {Yadlapalli}, {Yamaguchi},
  {Yfantis}, {Yoon}, {Young}, {Young}, {Younsi}, {Yu}, {Yuan}, {Yuan},
  {Zensus}, {Zhang}, {Zhao}, \& {Zhao}}]{EHTC+2023}
{Event Horizon Telescope Collaboration}, {Akiyama}, K., {Alberdi}, A., {et~al.}
  2023, \bibinfo{title}{{First M87 Event Horizon Telescope Results. IX.
  Detection of Near-horizon Circular Polarization},} \apjl, 957, L20,
  \dodoi{10.3847/2041-8213/acff70}

\bibitem[{ {Event Horizon Telescope Collaboration}
  {et~al.}(2024{\natexlab{a}}){Event Horizon Telescope Collaboration},
  {Akiyama}, {Alberdi}, {Alef}, {Algaba}, {Anantua}, {Asada}, {Azulay}, {Bach},
  {Baczko}, {Ball}, {Balokovic}, {Bandyopadhyay}, {Barrett}, {Baub{\"o}ck},
  {Benson}, {Bintley}, {Blackburn}, {Blundell}, {Bouman}, {Bower}, {Boyce},
  {Bremer}, {Brinkerink}, {Brissenden}, {Britzen}, {Broderick}, {Broguiere},
  {Bronzwaer}, {Bustamante}, {Byun}, {Carlstrom}, {Ceccobello}, {Chael},
  {Chan}, {Chang}, {Chatterjee}, {Chatterjee}, {Chen}, {Chen}, {Cheng}, {Cho},
  {Christian}, {Conroy}, {Conway}, {Cordes}, {Crawford}, {Crew}, {Cruz-Osorio},
  {Cui}, {Dahale}, {Davelaar}, {De Laurentis}, {Deane}, {Dempsey}, {Desvignes},
  {Dexter}, {Dhruv}, {Dihingia}, {Doeleman}, {Dougal}, {Dzib}, {Eatough},
  {Emami}, {Falcke}, {Farah}, {Fish}, {Fomalont}, {Ford}, {Foschi},
  {Fraga-Encinas}, {Freeman}, {Friberg}, {Fromm}, {Fuentes}, {Galison},
  {Gammie}, {Garc{\'\i}a}, {Gentaz}, {Georgiev}, {Goddi}, {Gold},
  {G{\'o}mez-Ruiz}, {G{\'o}mez}, {Gu}, {Gurwell}, {Hada}, {Haggard}, {Haworth},
  {Hecht}, {Hesper}, {Heumann}, {Ho}, {Ho}, {Honma}, {Huang}, {Huang},
  {Hughes}, {Ikeda}, {Impellizzeri}, {Inoue}, {Issaoun}, {James}, {Jannuzi},
  {Janssen}, {Jeter}, {Jiang}, {Jim{\'e}nez-Rosales}, {Johnson}, {Jorstad},
  {Joshi}, {Jung}, {Karami}, {Karuppusamy}, {Kawashima}, {Keating}, {Kettenis},
  {Kim}, {Kim}, {Kim}, {Kim}, {Kino}, {Koay}, {Kocherlakota}, {Kofuji}, {Koch},
  {Koyama}, {Kramer}, {Kramer}, {Kramer}, {Krichbaum}, {Kuo}, {La Bella},
  {Lauer}, {Lee}, {Lee}, {Leung}, {Levis}, {Li}, {Lico}, {Lindahl},
  {Lindqvist}, {Lisakov}, {Liu}, {Liu}, {Liuzzo}, {Lo}, {Lobanov}, {Loinard},
  {Lonsdale}, {Lowitz}, {Lu}, {MacDonald}, {Mao}, {Marchili}, {Markoff},
  {Marrone}, {Marscher}, {Mart{\'\i}-Vidal}, {Matsushita}, {Matthews},
  {Medeiros}, {Menten}, {Michalik}, {Mizuno}, {Mizuno}, {Moran}, {Moriyama},
  {Moscibrodzka}, {Mulaudzi}, {M{\"u}ller}, {M{\"u}ller}, {Mus}, {Musoke},
  {Myserlis}, {Nadolski}, {Nagai}, {Nagar}, {Nakamura}, {Narayanan},
  {Natarajan}, {Nathanail}, {Fuentes}, {Neilsen}, {Neri}, {Ni}, {Noutsos},
  {Nowak}, {Oh}, {Okino}, {Olivares}, {Ortiz-Le{\'o}n}, {Oyama}, {{\"O}zel},
  {Palumbo}, {Paraschos}, {Park}, {Parsons}, {Patel}, \& {Pen}}]{EHTC+2024b}
{Event Horizon Telescope Collaboration}, {Akiyama}, K., {Alberdi}, A., {et~al.}
  2024{\natexlab{a}}, \bibinfo{title}{{First Sagittarius A* Event Horizon
  Telescope Results. VII. Polarization of the Ring},} \apjl, 964, L25,
  \dodoi{10.3847/2041-8213/ad2df0}

\bibitem[{ {Event Horizon Telescope Collaboration}
  {et~al.}(2024{\natexlab{b}}){Event Horizon Telescope Collaboration},
  {Akiyama}, {Alberdi}, {Alef}, {Algaba}, {Anantua}, {Asada}, {Azulay}, {Bach},
  {Baczko}, {Ball}, {Balokovi{\'c}}, {Bandyopadhyay}, {Barrett}, {Baub{\"o}ck},
  {Benson}, {Bintley}, {Blackburn}, {Blundell}, {Bouman}, {Bower}, {Boyce},
  {Bremer}, {Brinkerink}, {Brissenden}, {Britzen}, {Broderick}, {Broguiere},
  {Bronzwaer}, {Bustamante}, {Byun}, {Carlstrom}, {Ceccobello}, {Chael},
  {Chan}, {Chang}, {Chatterjee}, {Chatterjee}, {Chen}, {Chen}, {Cheng}, {Cho},
  {Christian}, {Conroy}, {Conway}, {Cordes}, {Crawford}, {Crew}, {Cruz-Osorio},
  {Cui}, {Dahale}, {Davelaar}, {De Laurentis}, {Deane}, {Dempsey}, {Desvignes},
  {Dexter}, {Dhruv}, {Dihingia}, {Doeleman}, {Dougall}, {Dzib}, {Eatough},
  {Emami}, {Falcke}, {Farah}, {Fish}, {Fomalont}, {Ford}, {Foschi},
  {Fraga-Encinas}, {Freeman}, {Friberg}, {Fromm}, {Fuentes}, {Galison},
  {Gammie}, {Garc{\'\i}a}, {Gentaz}, {Georgiev}, {Goddi}, {Gold},
  {G{\'o}mez-Ruiz}, {G{\'o}mez}, {Gu}, {Gurwell}, {Hada}, {Haggard}, {Haworth},
  {Hecht}, {Hesper}, {Heumann}, {Ho}, {Ho}, {Honma}, {Huang}, {Huang},
  {Hughes}, {Ikeda}, {Impellizzeri}, {Inoue}, {Issaoun}, {James}, {Jannuzi},
  {Janssen}, {Jeter}, {Jiang}, {Jim{\'e}nez-Rosales}, {Johnson}, {Jorstad},
  {Joshi}, {Jung}, {Karami}, {Karuppusamy}, {Kawashima}, {Keating}, {Kettenis},
  {Kim}, {Kim}, {Kim}, {Kim}, {Kino}, {Koay}, {Kocherlakota}, {Kofuji}, {Koch},
  {Koyama}, {Kramer}, {Kramer}, {Kramer}, {Krichbaum}, {Kuo}, {La Bella},
  {Lauer}, {Lee}, {Lee}, {Leung}, {Levis}, {Li}, {Lico}, {Lindahl},
  {Lindqvist}, {Lisakov}, {Liu}, {Liu}, {Liuzzo}, {Lo}, {Lobanov}, {Loinard},
  {Lonsdale}, {Lowitz}, {Lu}, {MacDonald}, {Mao}, {Marchili}, {Markoff},
  {Marrone}, {Marscher}, {Mart{\'\i}-Vidal}, {Matsushita}, {Matthews},
  {Medeiros}, {Menten}, {Michalik}, {Mizuno}, {Mizuno}, {Moran}, {Moriyama},
  {Moscibrodzka}, {Mulaudzi}, {M{\"u}ller}, {M{\"u}ller}, {Mus}, {Musoke},
  {Myserlis}, {Nadolski}, {Nagai}, {Nagar}, {Nakamura}, {Narayanan},
  {Natarajan}, {Nathanail}, {Fuentes}, {Neilsen}, {Neri}, {Ni}, {Noutsos},
  {Nowak}, {Oh}, {Okino}, {Olivares}, {Ortiz-Le{\'o}n}, {Oyama}, {{\"O}zel},
  {Palumbo}, {Paraschos}, {Park}, {Parsons}, {Patel}, {Pen}, {Pesce},
  {Pi{\'e}tu}, {Plambeck}, {PopStefanija}, {Porth}, {P{\"o}tzl}, {Prather},
  {Preciado-L{\'o}pez}, {Psaltis}, {Pu}, {Ramakrishnan}, {Rao}, {Rawlings},
  {Raymond}, {Rezzolla}, {Ricarte}, {Ripperda}, {Roelofs}, {Rogers},
  {Romero-Ca{\~n}izales}, {Ros}, {Roshanineshat}, {Rottmann}, {Roy}, {Ruiz},
  {Ruszczyk}, {Rygl}, {S{\'a}nchez}, {S{\'a}nchez-Arg{\"u}elles},
  {S{\'a}nchez-Portal}, {Sasada}, {Satapathy}, {Savolainen}, {Schloerb},
  {Schonfeld}, {Schuster}, {Shao}, {Shen}, {Small}, {Sohn}, {SooHoo},
  {Sosapanta Salas}, {Souccar}, {Stanway}, {Sun}, {Tazaki}, {Tetarenko},
  {Tiede}, {Tilanus}, {Titus}, {Torne}, {Toscano}, {Traianou}, {Trent},
  {Trippe}, {Turk}, {van Bemmel}, {van Langevelde}, {van Rossum}, {Vos},
  {Wagner}, {Ward-Thompson}, {Wardle}, {Washington}, {Weintroub}, {Wharton},
  {Wielgus}, {Wiik}, {Witzel}, {Wondrak}, {Wong}, {Wu}, {Yadlapalli},
  {Yamaguchi}, {Yfantis}, {Yoon}, {Young}, {Young}, {Younsi}, {Yu}, {Yuan},
  {Yuan}, {Zensus}, {Zhang}, {Zhao}, {Zhao}, \& {Najafi-Ziyazi}}]{EHTC+2024c}
{Event Horizon Telescope Collaboration}, {Akiyama}, K., {Alberdi}, A., {et~al.}
  2024{\natexlab{b}}, \bibinfo{title}{{First Sagittarius A* Event Horizon
  Telescope Results. VIII. Physical Interpretation of the Polarized Ring},}
  \apjl, 964, L26, \dodoi{10.3847/2041-8213/ad2df1}

\bibitem[{V.~L. {Fish} {et~al.}(2009){Fish}, {Doeleman}, {Broderick}, {Loeb},
  \& {Rogers}}]{Fish+2009}
{Fish}, V.~L., {Doeleman}, S.~S., {Broderick}, A.~E., {Loeb}, A., \& {Rogers},
  A. E.~E. 2009, \bibinfo{title}{{Detecting Changing Polarization Structures in
  Sagittarius A* with High Frequency VLBI},} \apj, 706, 1353,
  \dodoi{10.1088/0004-637X/706/2/1353}

\bibitem[{L.~G. {Fishbone} \& V. {Moncrief}(1976){Fishbone} \&
  {Moncrief}}]{Fishbone&Moncrief1976}
{Fishbone}, L.~G., \& {Moncrief}, V. 1976, \bibinfo{title}{{Relativistic fluid
  disks in orbit around Kerr black holes.},} \apj, 207, 962,
  \dodoi{10.1086/154565}

\bibitem[{C.~F. {Gammie} {et~al.}(2003){Gammie}, {McKinney}, \&
  {T{\'o}th}}]{Gammie+2003}
{Gammie}, C.~F., {McKinney}, J.~C., \& {T{\'o}th}, G. 2003,
  \bibinfo{title}{{HARM: A Numerical Scheme for General Relativistic
  Magnetohydrodynamics},} \apj, 589, 444, \dodoi{10.1086/374594}

\bibitem[{Z. {Gelles} {et~al.}(2021){Gelles}, {Himwich}, {Johnson}, \&
  {Palumbo}}]{Gelles+2021}
{Gelles}, Z., {Himwich}, E., {Johnson}, M.~D., \& {Palumbo}, D. C.~M. 2021,
  \bibinfo{title}{{Polarized image of equatorial emission in the Kerr
  geometry},} \prd, 104, 044060, \dodoi{10.1103/PhysRevD.104.044060}

\bibitem[{B. {Georgiev} {et~al.}(2022){Georgiev}, {Pesce}, {Broderick}, {Wong},
  {Dhruv}, {Wielgus}, {Gammie}, {Chan}, {Chatterjee}, {Emami}, {Mizuno},
  {Gold}, {Fromm}, {Ricarte}, {Yoon}, {Joshi}, {Prather}, {Cruz-Osorio},
  {Johnson}, {Porth}, {Olivares}, {Younsi}, {Rezzolla}, {Vos}, {Qiu},
  {Nathanail}, {Narayan}, {Chael}, {Anantua}, {Moscibrodzka}, {Akiyama},
  {Alberdi}, {Alef}, {Algaba}, {Asada}, {Azulay}, {Bach}, {Baczko}, {Ball},
  {Balokovi{\'c}}, {Barrett}, {Baub{\"o}ck}, {Benson}, {Bintley}, {Blackburn},
  {Blundell}, {Bouman}, {Bower}, {Boyce}, {Bremer}, {Brinkerink}, {Brissenden},
  {Britzen}, {Broguiere}, {Bronzwaer}, {Bustamante}, {Byun}, {Carlstrom},
  {Ceccobello}, {Chatterjee}, {Chen}, {Chen}, {Cheng}, {Cho}, {Christian},
  {Conroy}, {Conway}, {Cordes}, {Crawford}, {Crew}, {Cui}, {Davelaar}, {De
  Laurentis}, {Deane}, {Dempsey}, {Desvignes}, {Dexter}, {Doeleman}, {Dougal},
  {Dzib}, {Eatough}, {Falcke}, {Farah}, {Fish}, {Fomalont}, {Ford},
  {Fraga-Encinas}, {Freeman}, {Friberg}, {Fuentes}, {Galison}, {Garc{\'\i}a},
  {Gentaz}, {Goddi}, {G{\'o}mez-Ruiz}, {G{\'o}mez}, {Gu}, {Gurwell}, {Hada},
  {Haggard}, {Haworth}, {Hecht}, {Hesper}, {Heumann}, {Ho}, {Ho}, {Honma},
  {Huang}, {Huang}, {Hughes}, {Ikeda}, {Impellizzeri}, {Inoue}, {Issaoun},
  {James}, {Jannuzi}, {Janssen}, {Jeter}, {Jiang}, {Jim{\'e}nez-Rosales},
  {Jorstad}, {Jung}, {Karami}, {Karuppusamy}, {Kawashima}, {Keating},
  {Kettenis}, {Kim}, {Kim}, {Kim}, {Kim}, {Kino}, {Koay}, {Kocherlakota},
  {Kofuji}, {Koch}, {Koyama}, {Kramer}, {Kramer}, {Krichbaum}, {Kuo}, {La
  Bella}, {Lauer}, {Lee}, {Lee}, {Lehner}, {Leung}, {Levis}, {Li}, {Lico},
  {Lindahl}, {Lindqvist}, {Lisakov}, {Liu}, {Liu}, {Liuzzo}, {Lo}, {Lobanov},
  {Loinard}, {Lonsdale}, {Lu}, {Mao}, {Marchili}, {Markoff}, {Marrone},
  {Marscher}, {Mart{\'\i}-Vidal}, {Matsushita}, {Matthews}, {Menten},
  {Michalik}, {Mizuno}, {Moran}, {Moriyama}, {M{\"u}ller}, {Mus}, {Musoke},
  {Myserlis}, {Nadolski}, {Nagai}, {Nagar}, {Nakamura}, {Narayanan},
  {Natarajan}, {Navarro Fuentes}, {Neilsen}, {Neri}, {Ni}, {Noutsos}, {Nowak},
  {Oh}, {Okino}, {Ortiz-Le{\'o}n}, {Oyama}, {Palumbo}, {Paraschos}, {Park},
  {Parsons}, {Patel}, \& {Pen}}]{Georgiev+2022}
{Georgiev}, B., {Pesce}, D.~W., {Broderick}, A.~E., {et~al.} 2022,
  \bibinfo{title}{{A Universal Power-law Prescription for Variability from
  Synthetic Images of Black Hole Accretion Flows},} \apjl, 930, L20,
  \dodoi{10.3847/2041-8213/ac65eb}

\bibitem[{A.~M. {Ghez} {et~al.}(2003){Ghez}, {Duch{\^e}ne}, {Matthews},
  {Hornstein}, {Tanner}, {Larkin}, {Morris}, {Becklin}, {Salim}, {Kremenek},
  {Thompson}, {Soifer}, {Neugebauer}, \& {McLean}}]{Ghez+2003}
{Ghez}, A.~M., {Duch{\^e}ne}, G., {Matthews}, K., {et~al.} 2003,
  \bibinfo{title}{{The First Measurement of Spectral Lines in a Short-Period
  Star Bound to the Galaxy's Central Black Hole: A Paradox of Youth},} \apjl,
  586, L127, \dodoi{10.1086/374804}

\bibitem[{A.~M. {Ghez} {et~al.}(2008){Ghez}, {Salim}, {Weinberg}, {Lu}, {Do},
  {Dunn}, {Matthews}, {Morris}, {Yelda}, {Becklin}, {Kremenek},
  {Milosavljevic}, \& {Naiman}}]{Ghez+2008}
{Ghez}, A.~M., {Salim}, S., {Weinberg}, N.~N., {et~al.} 2008,
  \bibinfo{title}{{Measuring Distance and Properties of the Milky Way's Central
  Supermassive Black Hole with Stellar Orbits},} \apj, 689, 1044,
  \dodoi{10.1086/592738}

\bibitem[{S. {Gillessen} {et~al.}(2017){Gillessen}, {Plewa}, {Eisenhauer},
  {Sari}, {Waisberg}, {Habibi}, {Pfuhl}, {George}, {Dexter}, {von Fellenberg},
  {Ott}, \& {Genzel}}]{Gillessen+2017}
{Gillessen}, S., {Plewa}, P.~M., {Eisenhauer}, F., {et~al.} 2017,
  \bibinfo{title}{{An Update on Monitoring Stellar Orbits in the Galactic
  Center},} \apj, 837, 30, \dodoi{10.3847/1538-4357/aa5c41}

\bibitem[{ {GRAVITY Collaboration} {et~al.}(2018){GRAVITY Collaboration},
  {Abuter}, {Amorim}, {Baub{\"o}ck}, {Berger}, {Bonnet}, {Brandner},
  {Cl{\'e}net}, {Coud{\'e} Du Foresto}, {de Zeeuw}, {Deen}, {Dexter}, {Duvert},
  {Eckart}, {Eisenhauer}, {F{\"o}rster Schreiber}, {Garcia}, {Gao}, {Gendron},
  {Genzel}, {Gillessen}, {Guajardo}, {Habibi}, {Haubois}, {Henning}, {Hippler},
  {Horrobin}, {Huber}, {Jim{\'e}nez-Rosales}, {Jocou}, {Kervella}, {Lacour},
  {Lapeyr{\`e}re}, {Lazareff}, {Le Bouquin}, {L{\'e}na}, {Lippa}, {Ott},
  {Panduro}, {Paumard}, {Perraut}, {Perrin}, {Pfuhl}, {Plewa}, {Rabien},
  {Rodr{\'\i}guez-Coira}, {Rousset}, {Sternberg}, {Straub}, {Straubmeier},
  {Sturm}, {Tacconi}, {Vincent}, {von Fellenberg}, {Waisberg}, {Widmann},
  {Wieprecht}, {Wiezorrek}, {Woillez}, \& {Yazici}}]{Gravity2018QU}
{GRAVITY Collaboration}, {Abuter}, R., {Amorim}, A., {et~al.} 2018,
  \bibinfo{title}{{Detection of orbital motions near the last stable circular
  orbit of the massive black hole SgrA*},} \aap, 618, L10,
  \dodoi{10.1051/0004-6361/201834294}

\bibitem[{ {GRAVITY Collaboration} {et~al.}(2020){GRAVITY Collaboration},
  {Jim{\'e}nez-Rosales}, {Dexter}, {Widmann}, {Baub{\"o}ck}, {Abuter},
  {Amorim}, {Berger}, {Bonnet}, {Brandner}, {Cl{\'e}net}, {de Zeeuw}, {Eckart},
  {Eisenhauer}, {F{\"o}rster Schreiber}, {Garcia}, {Gao}, {Gendron}, {Genzel},
  {Gillessen}, {Habibi}, {Haubois}, {Hei{\ss}el}, {Henning}, {Hippler},
  {Horrobin}, {Jochum}, {Jocou}, {Kaufer}, {Kervella}, {Lacour},
  {Lapeyr{\`e}re}, {Le Bouquin}, {L{\'e}na}, {Nowak}, {Ott}, {Paumard},
  {Perraut}, {Perrin}, {Pfuhl}, {Rodr{\'\i}guez-Coira}, {Shangguan},
  {Scheithauer}, {Stadler}, {Straub}, {Straubmeier}, {Sturm}, {Tacconi},
  {Vincent}, {von Fellenberg}, {Waisberg}, {Wieprecht}, {Wiezorrek}, {Woillez},
  {Yazici}, \& {Zins}}]{Gravity+2020b}
{GRAVITY Collaboration}, {Jim{\'e}nez-Rosales}, A., {Dexter}, J., {et~al.}
  2020, \bibinfo{title}{{Dynamically important magnetic fields near the event
  horizon of Sgr A*},} \aap, 643, A56, \dodoi{10.1051/0004-6361/202038283}

\bibitem[{ {GRAVITY Collaboration} {et~al.}(2022){GRAVITY Collaboration},
  {Abuter}, {Aimar}, {Amorim}, {Ball}, {Baub{\"o}ck}, {Berger}, {Bonnet},
  {Bourdarot}, {Brandner}, {Cardoso}, {Cl{\'e}net}, {Dallilar}, {Davies}, {de
  Zeeuw}, {Dexter}, {Drescher}, {Eisenhauer}, {F{\"o}rster Schreiber},
  {Foschi}, {Garcia}, {Gao}, {Gendron}, {Genzel}, {Gillessen}, {Habibi},
  {Haubois}, {Hei{\ss}el}, {Henning}, {Hippler}, {Horrobin}, {Jochum}, {Jocou},
  {Kaufer}, {Kervella}, {Lacour}, {Lapeyr{\`e}re}, {Le Bouquin}, {L{\'e}na},
  {Lutz}, {Ott}, {Paumard}, {Perraut}, {Perrin}, {Pfuhl}, {Rabien},
  {Shangguan}, {Shimizu}, {Scheithauer}, {Stadler}, {Stephens}, {Straub},
  {Straubmeier}, {Sturm}, {Tacconi}, {Tristram}, {Vincent}, {von Fellenberg},
  {Widmann}, {Wieprecht}, {Wiezorrek}, {Woillez}, {Yazici}, \&
  {Young}}]{Gravity+2022}
{GRAVITY Collaboration}, {Abuter}, R., {Aimar}, N., {et~al.} 2022,
  \bibinfo{title}{{Mass distribution in the Galactic Center based on
  interferometric astrometry of multiple stellar orbits},} \aap, 657, L12,
  \dodoi{10.1051/0004-6361/202142465}

\bibitem[{ {Gravity Collaboration} {et~al.}(2023){Gravity Collaboration},
  {Abuter}, {Aimar}, {Amaro Seoane}, {Amorim}, {Baub{\"o}ck}, {Berger},
  {Bonnet}, {Bourdarot}, {Brandner}, {Cardoso}, {Cl{\'e}net}, {Davies}, {de
  Zeeuw}, {Dexter}, {Drescher}, {Eckart}, {Eisenhauer}, {Feuchtgruber},
  {Finger}, {F{\"o}rster Schreiber}, {Foschi}, {Garcia}, {Gao}, {Gelles},
  {Gendron}, {Genzel}, {Gillessen}, {Hartl}, {Haubois}, {Haussmann},
  {Hei{\ss}el}, {Henning}, {Hippler}, {Horrobin}, {Jochum}, {Jocou}, {Kaufer},
  {Kervella}, {Lacour}, {Lapeyr{\`e}re}, {Le Bouquin}, {L{\'e}na}, {Lutz},
  {Mang}, {More}, {Ott}, {Paumard}, {Perraut}, {Perrin}, {Pfuhl}, {Rabien},
  {Ribeiro}, {Sadun Bordoni}, {Scheithauer}, {Shangguan}, {Shimizu}, {Stadler},
  {Straub}, {Straubmeier}, {Sturm}, {Tacconi}, {Vincent}, {von Fellenberg},
  {Widmann}, {Wielgus}, {Wieprecht}, {Wiezorrek}, \& {Woillez}}]{Gravity+2023}
{Gravity Collaboration}, {Abuter}, R., {Aimar}, N., {et~al.} 2023,
  \bibinfo{title}{{Polarimetry and astrometry of NIR flares as event horizon
  scale, dynamical probes for the mass of Sgr A*},} \aap, 677, L10,
  \dodoi{10.1051/0004-6361/202347416}

\bibitem[{A.~A. {Grigorian} \& J. {Dexter}(2024){Grigorian} \&
  {Dexter}}]{Grigorian&Dexter2024}
{Grigorian}, A.~A., \& {Dexter}, J. 2024, \bibinfo{title}{{The relationship
  between simulated sub-millimeter and near-infrared images of Sagittarius A*
  from a magnetically arrested black hole accretion flow},} \mnras, 530, 1563,
  \dodoi{10.1093/mnras/stae934}

\bibitem[{N. {Hamaus} {et~al.}(2009){Hamaus}, {Paumard}, {M{\"u}ller},
  {Gillessen}, {Eisenhauer}, {Trippe}, \& {Genzel}}]{Hamaus+2009}
{Hamaus}, N., {Paumard}, T., {M{\"u}ller}, T., {et~al.} 2009,
  \bibinfo{title}{{Prospects for Testing the Nature of Sgr A*'s Near-Infrared
  Flares on the Basis of Current Very Large Telescope{\textemdash}and Future
  Very Large Telescope Interferometer{\textemdash}Observations},} \apj, 692,
  902, \dodoi{10.1088/0004-637X/692/1/902}

\bibitem[{I.~V. {Igumenshchev} {et~al.}(2003){Igumenshchev}, {Narayan}, \&
  {Abramowicz}}]{Igumenshchev+2003}
{Igumenshchev}, I.~V., {Narayan}, R., \& {Abramowicz}, M.~A. 2003,
  \bibinfo{title}{{Three-dimensional Magnetohydrodynamic Simulations of
  Radiatively Inefficient Accretion Flows},} \apj, 592, 1042,
  \dodoi{10.1086/375769}

\bibitem[{S. {Issaoun} {et~al.}(2019){Issaoun}, {Johnson}, {Blackburn},
  {Brinkerink}, {Mo{\'s}cibrodzka}, {Chael}, {Goddi}, {Mart{\'\i}-Vidal},
  {Wagner}, {Doeleman}, {Falcke}, {Krichbaum}, {Akiyama}, {Bach}, {Bouman},
  {Bower}, {Broderick}, {Cho}, {Crew}, {Dexter}, {Fish}, {Gold}, {G{\'o}mez},
  {Hada}, {Hern{\'a}ndez-G{\'o}mez}, {Jan{\ss}en}, {Kino}, {Kramer}, {Loinard},
  {Lu}, {Markoff}, {Marrone}, {Matthews}, {Moran}, {M{\"u}ller}, {Roelofs},
  {Ros}, {Rottmann}, {Sanchez}, {Tilanus}, {de Vicente}, {Wielgus}, {Zensus},
  \& {Zhao}}]{Issaoun2019}
{Issaoun}, S., {Johnson}, M.~D., {Blackburn}, L., {et~al.} 2019,
  \bibinfo{title}{{The Size, Shape, and Scattering of Sagittarius A* at 86 GHz:
  First VLBI with ALMA},} \apj, 871, 30, \dodoi{10.3847/1538-4357/aaf732}

\bibitem[{H. {Jia} {et~al.}(2023){Jia}, {Ripperda}, {Quataert}, {White},
  {Chatterjee}, {Philippov}, \& {Liska}}]{Jia+2023}
{Jia}, H., {Ripperda}, B., {Quataert}, E., {et~al.} 2023,
  \bibinfo{title}{{Millimeter observational signatures of flares in
  magnetically arrested black hole accretion models},} \mnras, 526, 2924,
  \dodoi{10.1093/mnras/stad2935}

\bibitem[{A. {Jim{\'e}nez-Rosales} \& J. {Dexter}(2018){Jim{\'e}nez-Rosales} \&
  {Dexter}}]{Jimenez-Rosales&Dexter2018}
{Jim{\'e}nez-Rosales}, A., \& {Dexter}, J. 2018, \bibinfo{title}{{The impact of
  Faraday effects on polarized black hole images of Sagittarius A*},} \mnras,
  478, 1875, \dodoi{10.1093/mnras/sty1210}

\bibitem[{M.~D. {Johnson} {et~al.}(2015){Johnson}, {Loeb}, {Shiokawa}, {Chael},
  \& {Doeleman}}]{Johnson+2015}
{Johnson}, M.~D., {Loeb}, A., {Shiokawa}, H., {Chael}, A.~A., \& {Doeleman},
  S.~S. 2015, \bibinfo{title}{{Measuring the Direction and Angular Velocity of
  a Black Hole Accretion Disk via Lagged Interferometric Covariance},} \apj,
  813, 132, \dodoi{10.1088/0004-637X/813/2/132}

\bibitem[{A.~V. {Joshi} {et~al.}(2024){Joshi}, {Prather}, {Chan}, {Wielgus}, \&
  {Gammie}}]{Joshi+2024}
{Joshi}, A.~V., {Prather}, B.~S., {Chan}, C.-k., {Wielgus}, M., \& {Gammie},
  C.~F. 2024, \bibinfo{title}{{Circular Polarization of Simulated Images of
  Black Holes},} \apj, 972, 135, \dodoi{10.3847/1538-4357/ad5b51}

\bibitem[{J. {Knollm{\"u}ller} {et~al.}(2023){Knollm{\"u}ller}, {Arras}, \&
  {En{\ss}lin}}]{Knollmueller+2023}
{Knollm{\"u}ller}, J., {Arras}, P., \& {En{\ss}lin}, T. 2023,
  \bibinfo{title}{{Resolving Horizon-Scale Dynamics of Sagittarius A*},} arXiv
  e-prints, arXiv:2310.16889, \dodoi{10.48550/arXiv.2310.16889}

\bibitem[{A. {Levis} {et~al.}(2024){Levis}, {Chael}, {Bouman}, {Wielgus}, \&
  {Srinivasan}}]{Levis2024}
{Levis}, A., {Chael}, A.~A., {Bouman}, K.~L., {Wielgus}, M., \& {Srinivasan},
  P.~P. 2024, \bibinfo{title}{{Orbital polarimetric tomography of a flare near
  the Sagittarius A∗^{*} supermassive black hole},} Nature Astronomy, 8, 765,
  \dodoi{10.1038/s41550-024-02238-3}

\bibitem[{D.~P. {Marrone}(2006){Marrone}}]{Marrone2006}
{Marrone}, D.~P. 2006, PhD thesis, Harvard University

\bibitem[{D.~P. {Marrone} {et~al.}(2006){Marrone}, {Moran}, {Zhao}, \&
  {Rao}}]{Marrone+2006}
{Marrone}, D.~P., {Moran}, J.~M., {Zhao}, J.-H., \& {Rao}, R. 2006, in Journal
  of Physics Conference Series, Vol.~54, Journal of Physics Conference Series,
  ed. R.~{Sch{\"o}del}, G.~C. {Bower}, M.~P. {Muno}, S.~{Nayakshin}, \&
  T.~{Ott} (IOP), 354--362, \dodoi{10.1088/1742-6596/54/1/056}

\bibitem[{J.~M. {Michail} {et~al.}(2023){Michail}, {Yusef-Zadeh}, {Wardle}, \&
  {Kunneriath}}]{Michail+2023}
{Michail}, J.~M., {Yusef-Zadeh}, F., {Wardle}, M., \& {Kunneriath}, D. 2023,
  \bibinfo{title}{{Polarized signatures of adiabatically expanding hotspots in
  Sgr A*'s accretion flow},} \mnras, 520, 2644, \dodoi{10.1093/mnras/stad291}

\bibitem[{R.~S. {Millman} \& G.~D. {Parker}(1977){Millman} \&
  {Parker}}]{Millman&Parker1977}
{Millman}, R.~S., \& {Parker}, G.~D. 1977, {Elements of Differential Geometry}
  (Prentice Hall Inc.)

\bibitem[{M. {Mo{\'s}cibrodzka} {et~al.}(2017){Mo{\'s}cibrodzka}, {Dexter},
  {Davelaar}, \& {Falcke}}]{Moscibrodzka+2017}
{Mo{\'s}cibrodzka}, M., {Dexter}, J., {Davelaar}, J., \& {Falcke}, H. 2017,
  \bibinfo{title}{{Faraday rotation in GRMHD simulations of the jet launching
  zone of M87},} \mnras, 468, 2214, \dodoi{10.1093/mnras/stx587}

\bibitem[{M. {Mo{\'s}cibrodzka} {et~al.}(2016){Mo{\'s}cibrodzka}, {Falcke}, \&
  {Shiokawa}}]{Moscibrodzka+2016}
{Mo{\'s}cibrodzka}, M., {Falcke}, H., \& {Shiokawa}, H. 2016,
  \bibinfo{title}{{General relativistic magnetohydrodynamical simulations of
  the jet in M 87},} \aap, 586, A38, \dodoi{10.1051/0004-6361/201526630}

\bibitem[{M. {Mo{\'s}cibrodzka} \& C.~F. {Gammie}(2018){Mo{\'s}cibrodzka} \&
  {Gammie}}]{Moscibrodzka&Gammie2018}
{Mo{\'s}cibrodzka}, M., \& {Gammie}, C.~F. 2018, \bibinfo{title}{{IPOLE -
  semi-analytic scheme for relativistic polarized radiative transport},}
  \mnras, 475, 43, \dodoi{10.1093/mnras/stx3162}

\bibitem[{D.~J. {Mu{\~n}oz} {et~al.}(2012){Mu{\~n}oz}, {Marrone}, {Moran}, \&
  {Rao}}]{Munoz+2012}
{Mu{\~n}oz}, D.~J., {Marrone}, D.~P., {Moran}, J.~M., \& {Rao}, R. 2012,
  \bibinfo{title}{{The Circular Polarization of Sagittarius A* at Submillimeter
  Wavelengths},} \apj, 745, 115, \dodoi{10.1088/0004-637X/745/2/115}

\bibitem[{M. {Najafi-Ziyazi} {et~al.}(2024){Najafi-Ziyazi}, {Davelaar},
  {Mizuno}, \& {Porth}}]{Najafi-Ziyazi+2024}
{Najafi-Ziyazi}, M., {Davelaar}, J., {Mizuno}, Y., \& {Porth}, O. 2024,
  \bibinfo{title}{{Flares in the Galactic centre - II. Polarization signatures
  of flares at mm-wavelengths},} \mnras, 531, 3961,
  \dodoi{10.1093/mnras/stae1343}

\bibitem[{R. {Narayan} {et~al.}(2003){Narayan}, {Igumenshchev}, \&
  {Abramowicz}}]{Narayan+2003}
{Narayan}, R., {Igumenshchev}, I.~V., \& {Abramowicz}, M.~A. 2003,
  \bibinfo{title}{{Magnetically Arrested Disk: an Energetically Efficient
  Accretion Flow},} \pasj, 55, L69, \dodoi{10.1093/pasj/55.6.L69}

\bibitem[{R. {Narayan} {et~al.}(2012){Narayan}, {S{\"A} dowski}, {Penna}, \&
  {Kulkarni}}]{Narayan+2012}
{Narayan}, R., {S{\"A} dowski}, A., {Penna}, R.~F., \& {Kulkarni}, A.~K. 2012,
  \bibinfo{title}{{GRMHD simulations of magnetized advection-dominated
  accretion on a non-spinning black hole: role of outflows},} \mnras, 426,
  3241, \dodoi{10.1111/j.1365-2966.2012.22002.x}

\bibitem[{R. {Narayan} {et~al.}(2021){Narayan}, {Palumbo}, {Johnson}, {Gelles},
  {Himwich}, {Chang}, {Ricarte}, {Dexter}, {Gammie}, {Chael}, {Event Horizon
  Telescope Collaboration}, {Akiyama}, {Alberdi}, {Alef}, {Algaba}, {Anantua},
  {Asada}, {Azulay}, {Baczko}, {Ball}, {Balokovi{\'c}}, {Barrett}, {Benson},
  {Bintley}, {Blackburn}, {Blundell}, {Boland}, {Bouman}, {Bower}, {Boyce},
  {Bremer}, {Brinkerink}, {Brissenden}, {Britzen}, {Broderick}, {Broguiere},
  {Bronzwaer}, {Byun}, {Carlstrom}, {Chan}, {Chatterjee}, {Chatterjee}, {Chen},
  {Chen}, {Chesler}, {Cho}, {Christian}, {Conway}, {Cordes}, {Crawford},
  {Crew}, {Cruz-Osorio}, {Cui}, {Davelaar}, {De Laurentis}, {Deane}, {Dempsey},
  {Desvignes}, {Doeleman}, {Eatough}, {Falcke}, {Farah}, {Fish}, {Fomalont},
  {Ford}, {Fraga-Encinas}, {Friberg}, {Fromm}, {Fuentes}, {Galison},
  {Garc{\'\i}a}, {Gentaz}, {Georgiev}, {Goddi}, {Gold}, {G{\'o}mez},
  {G{\'o}mez-Ruiz}, {Gu}, {Gurwell}, {Hada}, {Haggard}, {Hecht}, {Hesper},
  {Ho}, {Ho}, {Honma}, {Huang}, {Huang}, {Hughes}, {Ikeda}, {Inoue}, {Issaoun},
  {James}, {Jannuzi}, {Janssen}, {Jeter}, {Jiang}, {Jimenez-Rosales},
  {Jorstad}, {Jung}, {Karami}, {Karuppusamy}, {Kawashima}, {Keating},
  {Kettenis}, {Kim}, {Kim}, {Kim}, {Kim}, {Kino}, {Koay}, {Kofuji}, {Koch},
  {Koyama}, {Kramer}, {Kramer}, {Krichbaum}, {Kuo}, {Lauer}, {Lee}, {Levis},
  {Li}, {Li}, {Lindqvist}, {Lico}, {Lindahl}, {Liu}, {Liu}, {Liuzzo}, {Lo},
  {Lobanov}, {Loinard}, {Lonsdale}, {Lu}, {MacDonald}, {Mao}, {Marchili},
  {Markoff}, {Marrone}, {Marscher}, {Mart{\'\i}-Vidal}, {Matsushita},
  {Matthews}, {Medeiros}, {Menten}, {Mizuno}, {Mizuno}, {Moran}, {Moriyama},
  {Moscibrodzka}, {M{\"u}ller}, {Musoke}, {Mej{\'\i}as}, {Nagai}, {Nagar},
  {Nakamura}, {Narayanan}, {Natarajan}, {Nathanail}, {Neilsen}, {Neri}, {Ni},
  {Noutsos}, {Nowak}, {Okino}, {Olivares}, {Ortiz-Le{\'o}n}, {Oyama},
  {{\"O}zel}, {Park}, {Patel}, {Pen}, {Pesce}, {Pi{\'e}tu}, {Plambeck},
  {PopStefanija}, {Porth}, {P{\"o}tzl}, {Prather}, {Preciado-L{\'o}pez},
  {Psaltis}, {Pu}, {Ramakrishnan}, {Rao}, {Rawlings}, {Raymond}, {Rezzolla},
  {Ripperda}, {Roelofs}, {Rogers}, {Ros}, {Rose}, {Roshanineshat}, {Rottmann},
  {Roy}, {Ruszczyk}, {Rygl}, {S{\'a}nchez}, {S{\'a}nchez-Arguelles}, \&
  {Sasada}}]{Narayan+2021}
{Narayan}, R., {Palumbo}, D. C.~M., {Johnson}, M.~D., {et~al.} 2021,
  \bibinfo{title}{{The Polarized Image of a Synchrotron-emitting Ring of Gas
  Orbiting a Black Hole},} \apj, 912, 35, \dodoi{10.3847/1538-4357/abf117}

\bibitem[{H.~R. {Olivares} {et~al.}(2023){Olivares}, {Mo{\'s}cibrodzka}, \&
  {Porth}}]{Olivares+2023}
{Olivares}, H.~R., {Mo{\'s}cibrodzka}, M.~A., \& {Porth}, O. 2023,
  \bibinfo{title}{{General relativistic hydrodynamic simulations of perturbed
  transonic accretion},} \aap, 678, A141, \dodoi{10.1051/0004-6361/202346010}

\bibitem[{D.~C.~M. {Palumbo} {et~al.}(2022){Palumbo}, {Gelles}, {Tiede},
  {Chang}, {Pesce}, {Chael}, \& {Johnson}}]{Palumbo_2022}
{Palumbo}, D. C.~M., {Gelles}, Z., {Tiede}, P., {et~al.} 2022,
  \bibinfo{title}{{Bayesian Accretion Modeling: Axisymmetric Equatorial
  Emission in the Kerr Spacetime},} \apj, 939, 107,
  \dodoi{10.3847/1538-4357/ac9ab7}

\bibitem[{D.~C.~M. {Palumbo} {et~al.}(2020){Palumbo}, {Wong}, \&
  {Prather}}]{Palumbo+2020}
{Palumbo}, D. C.~M., {Wong}, G.~N., \& {Prather}, B.~S. 2020,
  \bibinfo{title}{{Discriminating Accretion States via Rotational Symmetry in
  Simulated Polarimetric Images of M87},} arXiv e-prints, arXiv:2004.01751.
\newblock \doarXiv{2004.01751}

\bibitem[{B.~S. {Prather}(2024){Prather}}]{Prather2024}
{Prather}, B.~S. 2024, \bibinfo{title}{{KHARMA: Flexible, Portable Performance
  for GRMHD},} arXiv e-prints, arXiv:2408.01361,
  \dodoi{10.48550/arXiv.2408.01361}

\bibitem[{R. {Qiu} {et~al.}(2023){Qiu}, {Ricarte}, {Narayan}, {Wong}, {Chael},
  \& {Palumbo}}]{Qiu+2023}
{Qiu}, R., {Ricarte}, A., {Narayan}, R., {et~al.} 2023, \bibinfo{title}{{Using
  Machine Learning to link black hole accretion flows with spatially resolved
  polarimetric observables},} \mnras, 520, 4867, \dodoi{10.1093/mnras/stad466}

\bibitem[{S.~M. {Ressler} {et~al.}(2023){Ressler}, {White}, \&
  {Quataert}}]{Ressler+2023}
{Ressler}, S.~M., {White}, C.~J., \& {Quataert}, E. 2023,
  \bibinfo{title}{{Wind-fed GRMHD simulations of Sagittarius A*: tilt and
  alignment of jets and accretion discs, electron thermodynamics, and
  multiscale modelling of the rotation measure},} \mnras, 521, 4277,
  \dodoi{10.1093/mnras/stad837}

\bibitem[{S.~M. {Ressler} {et~al.}(2020){Ressler}, {White}, {Quataert}, \&
  {Stone}}]{Ressler+2020}
{Ressler}, S.~M., {White}, C.~J., {Quataert}, E., \& {Stone}, J.~M. 2020,
  \bibinfo{title}{{Ab Initio Horizon-scale Simulations of Magnetically Arrested
  Accretion in Sagittarius A* Fed by Stellar Winds},} \apjl, 896, L6,
  \dodoi{10.3847/2041-8213/ab9532}

\bibitem[{A. {Ricarte} {et~al.}(2023){Ricarte}, {Johnson}, {Kovalev},
  {Palumbo}, \& {Emami}}]{Ricarte+2023c}
{Ricarte}, A., {Johnson}, M.~D., {Kovalev}, Y.~Y., {Palumbo}, D. C.~M., \&
  {Emami}, R. 2023, \bibinfo{title}{{How Spatially Resolved Polarimetry Informs
  Black Hole Accretion Flow Models},} Galaxies, 11, 5,
  \dodoi{10.3390/galaxies11010005}

\bibitem[{A. {Ricarte} {et~al.}(2025){Ricarte}, {Natarajan}, {Narayan}, \&
  {Palumbo}}]{Ricarte+2025}
{Ricarte}, A., {Natarajan}, P., {Narayan}, R., \& {Palumbo}, D. C.~M. 2025,
  \bibinfo{title}{{Multimessenger Probes of Supermassive Black Hole Spin
  Evolution},} \apj, 980, 136, \dodoi{10.3847/1538-4357/ad9ea9}

\bibitem[{A. {Ricarte} {et~al.}(2020){Ricarte}, {Prather}, {Wong}, {Narayan},
  {Gammie}, \& {Johnson}}]{Ricarte+2020}
{Ricarte}, A., {Prather}, B.~S., {Wong}, G.~N., {et~al.} 2020,
  \bibinfo{title}{{Decomposing the Internal Faraday Rotation of Black Hole
  Accretion Flows},} \mnras, \dodoi{10.1093/mnras/staa2692}

\bibitem[{A. {Ricarte} {et~al.}(2021){Ricarte}, {Qiu}, \&
  {Narayan}}]{Ricarte+2021b}
{Ricarte}, A., {Qiu}, R., \& {Narayan}, R. 2021, \bibinfo{title}{{Black hole
  magnetic fields and their imprint on circular polarization images},} \mnras,
  505, 523, \dodoi{10.1093/mnras/stab1289}

\bibitem[{B. {Ripperda} {et~al.}(2022){Ripperda}, {Liska}, {Chatterjee},
  {Musoke}, {Philippov}, {Markoff}, {Tchekhovskoy}, \&
  {Younsi}}]{Ripperda+2022}
{Ripperda}, B., {Liska}, M., {Chatterjee}, K., {et~al.} 2022,
  \bibinfo{title}{{Black Hole Flares: Ejection of Accreted Magnetic Flux
  through 3D Plasmoid-mediated Reconnection},} \apjl, 924, L32,
  \dodoi{10.3847/2041-8213/ac46a1}

\bibitem[{R. {Sch{\"o}del} {et~al.}(2002){Sch{\"o}del}, {Ott}, {Genzel},
  {Hofmann}, {Lehnert}, {Eckart}, {Mouawad}, {Alexander}, {Reid}, {Lenzen},
  {Hartung}, {Lacombe}, {Rouan}, {Gendron}, {Rousset}, {Lagrange}, {Brandner},
  {Ageorges}, {Lidman}, {Moorwood}, {Spyromilio}, {Hubin}, \&
  {Menten}}]{Schoedel+2002}
{Sch{\"o}del}, R., {Ott}, T., {Genzel}, R., {et~al.} 2002, \bibinfo{title}{{A
  star in a 15.2-year orbit around the supermassive black hole at the centre of
  the Milky Way},} \nat, 419, 694, \dodoi{10.1038/nature01121}

\bibitem[{A. {S{\k{a}}dowski} {et~al.}(2013){S{\k{a}}dowski}, {Narayan},
  {Penna}, \& {Zhu}}]{Sadowski+2013}
{S{\k{a}}dowski}, A., {Narayan}, R., {Penna}, R., \& {Zhu}, Y. 2013,
  \bibinfo{title}{{Energy, momentum and mass outflows and feedback from thick
  accretion discs around rotating black holes},} \mnras, 436, 3856,
  \dodoi{10.1093/mnras/stt1881}

\bibitem[{A. {Tchekhovskoy} {et~al.}(2011){Tchekhovskoy}, {Narayan}, \&
  {McKinney}}]{Tchekhovskoy+2011}
{Tchekhovskoy}, A., {Narayan}, R., \& {McKinney}, J.~C. 2011,
  \bibinfo{title}{{Efficient generation of jets from magnetically arrested
  accretion on a rapidly spinning black hole},} \mnras, 418, L79,
  \dodoi{10.1111/j.1745-3933.2011.01147.x}

\bibitem[{Y. {Tsunetoe} {et~al.}(2021){Tsunetoe}, {Mineshige}, {Ohsuga},
  {Kawashima}, \& {Akiyama}}]{Tsunetoe+2021}
{Tsunetoe}, Y., {Mineshige}, S., {Ohsuga}, K., {Kawashima}, T., \& {Akiyama},
  K. 2021, \bibinfo{title}{{Polarization images of accretion flow around
  supermassive black holes: Imprints of toroidal field structure},} \pasj, 73,
  912, \dodoi{10.1093/pasj/psab054}

\bibitem[{F.~H. {Vincent} {et~al.}(2024){Vincent}, {Wielgus}, {Aimar},
  {Paumard}, \& {Perrin}}]{Vincent2024}
{Vincent}, F.~H., {Wielgus}, M., {Aimar}, N., {Paumard}, T., \& {Perrin}, G.
  2024, \bibinfo{title}{{Polarized signatures of orbiting hot spots: Special
  relativity impact and probe of spacetime curvature},} \aap, 684, A194,
  \dodoi{10.1051/0004-6361/202348016}

\bibitem[{J. {Vos} {et~al.}(2022){Vos}, {Mo{\'s}cibrodzka}, \&
  {Wielgus}}]{Vos2022}
{Vos}, J., {Mo{\'s}cibrodzka}, M.~A., \& {Wielgus}, M. 2022,
  \bibinfo{title}{{Polarimetric signatures of hot spots in black hole accretion
  flows},} \aap, 668, A185, \dodoi{10.1051/0004-6361/202244840}

\bibitem[{Y. {Wang} \& B. {Zhang}(2024){Wang} \& {Zhang}}]{Wang&Zhang2024}
{Wang}, Y., \& {Zhang}, B. 2024, \bibinfo{title}{{Evidence of a past merger of
  the Galactic Centre black hole},} Nature Astronomy, 8, 1592,
  \dodoi{10.1038/s41550-024-02358-w}

\bibitem[{M. {Wielgus} {et~al.}(2024){Wielgus}, {Issaoun}, {Mart{\'\i}-Vidal},
  {Emami}, {Moscibrodzka}, {Brinkerink}, {Goddi}, \& {Fomalont}}]{Wielgus+2024}
{Wielgus}, M., {Issaoun}, S., {Mart{\'\i}-Vidal}, I., {et~al.} 2024,
  \bibinfo{title}{{The internal Faraday screen of Sagittarius A*},} \aap, 682,
  A97, \dodoi{10.1051/0004-6361/202347772}

\bibitem[{M. {Wielgus} {et~al.}(2022{\natexlab{a}}){Wielgus}, {Marchili},
  {Mart{\'\i}-Vidal}, {Keating}, {Ramakrishnan}, {Tiede}, {Fomalont},
  {Issaoun}, {Neilsen}, {Nowak}, {Blackburn}, {Gammie}, {Goddi}, {Haggard},
  {Lee}, {Moscibrodzka}, {Tetarenko}, {Bower}, {Chan}, {Chatterjee}, {Chesler},
  {Dexter}, {Doeleman}, {Georgiev}, {Gurwell}, {Johnson}, {Marrone}, {Mus},
  {Psaltis}, {Ripperda}, {Witzel}, {Akiyama}, {Alberdi}, {Alef}, {Carlos
  Algaba}, {Anantua}, {Asada}, {Azulay}, {Bach}, {Baczko}, {Ball},
  {Balokovi{\'c}}, {Barrett}, {Baub{\"o}ck}, {Benson}, {Bintley}, {Blundell},
  {Boland}, {Bouman}, {Boyce}, {Bremer}, {Brinkerink}, {Brissenden}, {Britzen},
  {Broderick}, {Broguiere}, {Bronzwaer}, {Bustamante}, {Byun}, {Carlstrom},
  {Ceccobello}, {Chael}, {Chatterjee}, {Chen}, {Chen}, {Cho}, {Christian},
  {Conroy}, {Conway}, {Cordes}, {Crawford}, {Crew}, {Cruz-Osorio}, {Cui},
  {Davelaar}, {De Laurentis}, {Deane}, {Dempsey}, {Desvignes}, {Dhruv}, {Dzib},
  {Eatough}, {Emami}, {Falcke}, {Farah}, {Fish}, {Alyson Ford},
  {Fraga-Encinas}, {Freeman}, {Friberg}, {Fromm}, {Fuentes}, {Galison},
  {Garc{\'\i}a}, {Gentaz}, {Gold}, {G{\'o}mez-Ruiz}, {G{\'o}mez}, {Gu}, {Hada},
  {Haworth}, {Hecht}, {Hesper}, {Ho}, {Ho}, {Honma}, {Huang}, {Huang},
  {Hughes}, {Ikeda}, {Violette Impellizzeri}, {Inoue}, {James}, {Jannuzi},
  {Janssen}, {Jeter}, {Jiang}, {Jim{\'e}nez-Rosales}, {Jorstad}, {Joshi},
  {Jung}, {Karami}, {Karuppusamy}, {Kawashima}, {Kettenis}, {Kim}, {Kim},
  {Kim}, {Kim}, {Kino}, {Yi Koay}, {Kocherlakota}, {Kofuji}, {Koch}, {Koyama},
  {Kramer}, {Kramer}, {Krichbaum}, {Kuo}, {La Bella}, {Lauer}, {Lee}, {Kin
  Leung}, {Levis}, {Li}, {Lico}, {Lindahl}, {Lindqvist}, {Lisakov}, {Liu},
  {Liu}, {Liuzzo}, {Lo}, {Lobanov}, {Loinard}, {Lonsdale}, {Lu}, {Mao},
  {Markoff}, {Marscher}, {Matsushita}, {Matthews}, {Medeiros}, {Menten},
  {Michalik}, {Mizuno}, {Mizuno}, {Moran}, {Moriyama}, {M{\"u}ller}, {Musoke},
  {Myserlis}, {Nadolski}, {Nagai}, {Nagar}, {Nakamura}, {Narayan}, {Narayanan},
  {Natarajan}, {Nathanail}, {Navarro Fuentes}, {Neri}, {Ni}, {Noutsos}, {Oh},
  {Okino}, {Olivares}, {Ortiz-Le{\'o}n}, {Oyama}, {{\"O}zel}, {Palumbo},
  {Filippos Paraschos}, {Park}, {Parsons}, {Patel}, {Pen}, {Pesce},
  {Pi{\'e}tu}, {Plambeck}, {PopStefanija}, {Porth}, {P{\"o}tzl}, {Prather},
  {Preciado-L{\'o}pez}, {Pu}, {Rao}, {Rawlings}, {Raymond}, {Rezzolla},
  {Ricarte}, {Roelofs}, {Rogers}, {Ros}, {Romero-Canizales}, {Roshanineshat},
  {Rottmann}, {Roy}, {Ruiz}, {Ruszczyk}, {Rygl}, {S{\'a}nchez},
  {S{\'a}nchez-Arg{\"u}elles}, {S{\'a}nchez-Portal}, {Sasada}, {Satapathy},
  {Savolainen}, {Peter Schloerb}, {Schuster}, {Shao}, {Shen}, {Small}, {Won
  Sohn}, {SooHoo}, {Souccar}, {Sun}, {Tazaki}, {Tilanus}, {Titus}, {Torne},
  {Traianou}, {Trent}, {Trippe}, {van Bemmel}, {Jan van Langevelde}, {van
  Rossum}, {Vos}, {Wagner}, {Ward-Thompson}, {Wardle}, {Weintroub}, {Wex},
  {Wharton}, {Wiik}, {Wondrak}, {Wong}, {Wu}, {Yamaguchi}, {Yoon}, {Young},
  {Young}, {Younsi}, {Yuan}, {Yuan}, {Anton Zensus}, {Zhang}, {Zhao}, \&
  {Zhao}}]{Wielgus+2022}
{Wielgus}, M., {Marchili}, N., {Mart{\'\i}-Vidal}, I., {et~al.}
  2022{\natexlab{a}}, \bibinfo{title}{{Millimeter Light Curves of Sagittarius
  A* Observed during the 2017 Event Horizon Telescope Campaign},} \apjl, 930,
  L19, \dodoi{10.3847/2041-8213/ac6428}

\bibitem[{M. {Wielgus} {et~al.}(2022{\natexlab{b}}){Wielgus}, {Moscibrodzka},
  {Vos}, {Gelles}, {Mart{\'\i}-Vidal}, {Farah}, {Marchili}, {Goddi}, \&
  {Messias}}]{Wielgus2022QU}
{Wielgus}, M., {Moscibrodzka}, M., {Vos}, J., {et~al.} 2022{\natexlab{b}},
  \bibinfo{title}{{Orbital motion near Sagittarius A$^{*}$ . Constraints from
  polarimetric ALMA observations},} \aap, 665, L6,
  \dodoi{10.1051/0004-6361/202244493}

\bibitem[{G.~N. {Wong} {et~al.}(2022){Wong}, {Prather}, {Dhruv}, {Ryan},
  {Mo{\'s}cibrodzka}, {Chan}, {Joshi}, {Yarza}, {Ricarte}, {Shiokawa},
  {Dolence}, {Noble}, {McKinney}, \& {Gammie}}]{Wong+2022}
{Wong}, G.~N., {Prather}, B.~S., {Dhruv}, V., {et~al.} 2022,
  \bibinfo{title}{{PATOKA: Simulating Electromagnetic Observables of Black Hole
  Accretion},} \apjs, 259, 64, \dodoi{10.3847/1538-4365/ac582e}

\bibitem[{A.~I. {Yfantis} {et~al.}(2024{\natexlab{a}}){Yfantis},
  {Mo{\'s}cibrodzka}, {Wielgus}, {Vos}, \& {Jimenez-Rosales}}]{Yfantis2024}
{Yfantis}, A.~I., {Mo{\'s}cibrodzka}, M.~A., {Wielgus}, M., {Vos}, J.~T., \&
  {Jimenez-Rosales}, A. 2024{\natexlab{a}}, \bibinfo{title}{{Fitting the light
  curves of Sagittarius A* with a hot-spot model. Bayesian modeling of QU loops
  in the millimeter band},} \aap, 685, A142,
  \dodoi{10.1051/0004-6361/202348230}

\bibitem[{A.~I. {Yfantis} {et~al.}(2024{\natexlab{b}}){Yfantis}, {Wielgus}, \&
  {Mo{\'s}cibrodzka}}]{Yfantis+2024b}
{Yfantis}, A.~I., {Wielgus}, M., \& {Mo{\'s}cibrodzka}, M. 2024{\natexlab{b}},
  \bibinfo{title}{{Hot spots around Sagittarius A*: Joint fits to astrometry
  and polarimetry},} \aap, 691, A327, \dodoi{10.1051/0004-6361/202451884}

\bibitem[{F. {Yuan} \& R. {Narayan}(2014){Yuan} \&
  {Narayan}}]{Yuan&Narayan2014}
{Yuan}, F., \& {Narayan}, R. 2014, \bibinfo{title}{{Hot Accretion Flows Around
  Black Holes},} \araa, 52, 529, \dodoi{10.1146/annurev-astro-082812-141003}

\end{thebibliography}

\end{document}